\documentclass[aip,apl,amsmath,amssymb,reprint]{revtex4-1}

\usepackage{graphicx}
\usepackage{dcolumn}
\usepackage{bm}

\usepackage[utf8]{inputenc}
\usepackage[T1]{fontenc}
\usepackage{mathptmx}
\usepackage{etoolbox}

\usepackage{physics}
\usepackage{xcolor}
\usepackage{mathtools}
\usepackage[hidelinks]{hyperref}

\newcommand{\rev}[1]{{\color{black} #1}} 

\makeatletter
\def\@email#1#2{%
 \endgroup
 \patchcmd{\titleblock@produce}
  {\frontmatter@RRAPformat}
  {\frontmatter@RRAPformat{\produce@RRAP{*#1\href{mailto:#2}{#2}}}\frontmatter@RRAPformat}
  {}{}
}%
\makeatother
\begin{document}


\title[Robust Wave Splitters Based on Scattering Singularities in Complex non-Hermitian Systems]{Robust Wave Splitters Based on Scattering Singularities in Complex non-Hermitian Systems}
\author{Jared Erb}
\email{jmerb@umd.edu}
\affiliation{Maryland Quantum Materials Center, Department of Physics, University of Maryland, College Park, MD, 20742, USA}

\author{Nadav Shaibe}%
\affiliation{Maryland Quantum Materials Center, Department of Physics, University of Maryland, College Park, MD, 20742, USA}

\author{Tsampikos Kottos}
\affiliation{Wave Transport in Complex Systems Lab, Physics Department, Wesleyan University, Middletown, CT, 06459, USA}

\author{Steven M. Anlage}
\affiliation{Maryland Quantum Materials Center, Department of Physics, University of Maryland, College Park, MD, 20742, USA}

\date{\today}
\begin{abstract}
\rev{We have discovered specific conditions for generic scattering systems to act as wave splitters that are robust to any change in relative amplitude or phase of an arbitrary injected waveform. Specifically for complex systems with tunable parameters, these conditions for robust splitting (RS) are abundant, and by using multiple tunable parameters the relative amplitude and phase of the output signals can also be tuned. The splitting property of the systems works for all possible input phase differences and amplitude ratios and does not require a particular coherent input signal.} We show experimentally that the fixed splitting ratios and output phases at \rev{RS conditions} are robust to 100 dB of relative power and 2$\pi$ phase changes of the input waves to a complex non-Hermitian two-port system. We also demonstrate that the splitting power ratio can be tuned by multiple orders of magnitude and the \rev{RS conditions} can be tuned to any desired frequency with suitable tunable perturbations embedded in the system. Although this phenomenon is realized in two-port systems \rev{and involves some degree of attenuation}, tunable robust splitting can be achieved between any two ports of multiport systems. These results are general to all wave scattering phenomena \rev{(electromagnetic, acoustic, etc.)} and hold in generic complex scattering systems.
\end{abstract}

\maketitle


Coherent Perfect Absorption (CPA) is a remarkable phenomenon where in a lossy system, injected electromagnetic radiation is completely absorbed, such that no incident energy leaves the system through reflection or transmission. This phenomenon was first discovered in the context of time-reversed lasers \cite{Chong10,Wan11,Baranov17}. Under this condition, a system acts as a lossy resonant cavity where a coherent excitation, corresponding to the scattering matrix eigenvector associated with the zero eigenvalue, results in perfect absorption. This has attracted much attention for its potential applications, including sensing \cite{Gupta12,Zhang14,Meng17}, photodetection \cite{Roger15,Vetlugin21,Vetlugin23}, filtering \cite{Zhang12,Faul25}, communication \cite{Bruck13,Sun14,Wong16,Imani20,Hougne21}, wireless power transfer \cite{Krasnok18}, wavefront shaping \cite{Pichler19,Chen20}, broadband absorption \cite{Pu12,Song14,Suwunnarat22}, etc. These applications can be implemented in any wave-scattering system, as demonstrated by studies of CPA using Random Matrix Theory (RMT) models of generic scattering systems \cite{Li17,Fyodorov17}. 

In the past, CPA was most commonly demonstrated in systems that are highly symmetric. In contrast, more recent work has shown that generic multi-modal cavities with or without reciprocity can be manipulated to create the conditions needed for CPA's at arbitrary frequencies \cite{Pichler19,Imani20,Chen20,Frazier20,Hougne21,Hougne21_2,Chen21,Erb24,Wang24,Shaibe25,Erb25}. \rev{Inspired by foundational work on CPA, and newer work expanding applications of CPA to generic complex scattering systems, we now recognize that the conditions required to display CPA also present an opportunity for another phenomenon, which we term ``robust splitting" (RS). We demonstrate that at these conditions, any signal sent into the system will be output with a particular fixed relative amplitude and phase (albeit with some overall attenuation) which depend on system-specific details. In the limiting case that the CPA waveform is injected, then the signal is perfectly absorbed.} 

One of the key \rev{features} in the newer works \rev{on creating the conditions required for CPA,} compared to previous ones, is the implementation of electronically-controlled tunable parameters embedded within a system. These tunable parameters allow for a more thorough manipulation of the scattering singularities of a system, including \rev{RS conditions}, and their dynamics. Due to the topological protection of the scattering singularities, along with the ease of finding and manipulating them with tunable parameters, the potential to find new applications of these singularities is greatly increased. 

In this paper, we show experimentally and theoretically that for all arbitrary monochromatic signals injected into a two-port system set to \rev{RS} conditions (excluding the \rev{CPA waveform} corresponding to the zero eigenvalue), \rev{using one port or both ports simultaneously}, the output signals will have a robust relative amplitude and phase. Additionally, due to the topological stability and manipulability of \rev{RS} conditions, the values of the relative amplitude and phase can be continuously varied by tuning a third parameter of the system. We also demonstrate the combined case of \rev{RS} conditions and a \rev{scattering matrix} exceptional point degeneracy in a non-reciprocal system for the first time.

To demonstrate the generality of \rev{robust splitting} conditions, we employ complex multi-modal physical systems interrogated by $M=2$ scattering channels. These systems are characterized by a $2\times 2$ scattering matrix $S$ which relates the incoming and outgoing wave excitations for each channel:

\begin{equation}
\begin{pmatrix}
V_1^{\rm out}\\
V_2^{\rm out}
\end{pmatrix}
 = S
 \begin{pmatrix}
V_1^{\rm in}\\
V_2^{\rm in}
\end{pmatrix}, \label{S_def}
\end{equation}

where $V_n^{\rm in},V_n^{\rm out}$ are the complex voltages of the incoming and outgoing excitations on channel $n=1,2$ respectively. The scattering matrix is experimentally measured in the frequency domain using a Keysight model N5242B microwave vector network analyzer (VNA) connected to two channels of the systems using coaxial cables. Due to the lossiness of the physical systems, $S$ is sub-unitary, and due to the underlying complex (chaotic) ray dynamics its complex matrix elements are irregular functions of frequency. 

\begin{figure*}[htb]
\centering
\includegraphics[width=18cm]{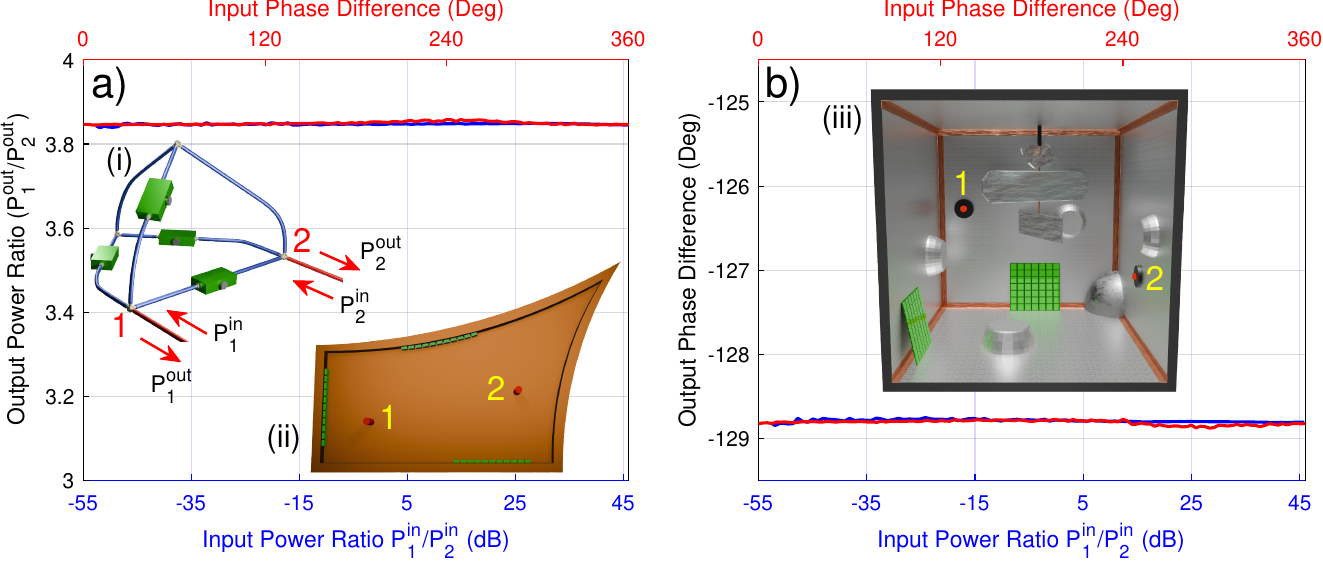}
\caption{Experimental demonstration of the robust splitting conditions in the quarter bow-tie two-dimensional billiard and schematics of three experimental systems. (a) Output power ratio vs input power ratio (lower axis, blue) and input phase difference (upper axis, red). The blue curve corresponds to the input signals power ratio being swept at a fixed phase difference, and the red curve corresponds to the input signals relative phase being swept at a fixed power ratio. Left inset (i) depicts a schematic view of a quasi-one-dimensional tetrahedral microwave graph with phase shifters ($TM^{0D}_{p=1,2,3,4}$) along four of the bonds. The ports connected to the graph are indicated by the red cables and numbers. The arrows indicate the input and output signals to/from the graph. Right inset (ii) depicts a schematic view of a quarter bow-tie two-dimensional billiard with three one-dimensional metasurfaces ($TM^{1D}_{p=1,2,3}$) along the walls of the billiard. The ports connected to the billiard are indicated by the red cylinders and yellow numbers. (b) Output phase difference vs input power ratio (lower axis, blue) and input phase difference (upper axis, red). The blue and red curves have the same interpretation as in (a). Inset (iii) depicts a schematic view of a three-dimensional microwave cavity with two two-dimensional metasurfaces ($TM^{2D}_{p=1,2}$) along the walls of the cavity. The ports connected to the cavity are indicated by the red cylinders and yellow numbers.}
\label{Robust_CPA}
\end{figure*}

The systems used in this work are quasi-one-dimensional microwave graphs (reciprocal and non-reciprocal) \cite{Kottos97,Kottos00,Kottos03,Hul04,Lawniczak10}, a quasi-two-dimensional microwave billiard \cite{Stockmann90,Doron90,So95,Gokir98,Hlushchuk00,Dietz07}, and a three-dimensional microwave cavity \cite{Deus95,Alt97,Frazier20,Frazier22} (see insets (i-iii) of Figure \ref{Robust_CPA}). To exert control over the scattering properties, each system contains electronically-controlled metasurfaces or phase-shifters which modify the amplitude and phase of reflecting and/or propagating waves. These tunable devices are denoted as $TM_p^{qD}$, where $q$ indicates the dimension of the device ($q=0$ for phase shifters and $q=1,2$ for the one and two-dimensional metasurfaces) and $p=1,2,3,...$ labels each device within a system. \rev{The number of tunable devices in each experimental system was determined by the number available, but in general two tunable devices are enough to achieve all the results shown in this paper. However, the more tunable devices in use, the easier it is to accomplish any specific condition.} The metasurfaces used in this work are tuned by a global voltage bias which modulates the amplitude and phase of reflected waves \cite{PhysRevApplied.20.014004,Erb24,Erb25}. For additional details on the \rev{measurements and} experimental systems used, see supplementary material \ref{sec.Exp} and Ref. \citenum{Erb25}. Embedding tunable devices in the systems allows for multi-dimensional parametric variation of the scattering matrix and its singularities, including robust splitting conditions. This capability enables us to find numerous scattering singularities which we can manipulate and utilize for applications.

To identify scattering singularities in a multi-dimensional parameter space, we use conditions of the scattering matrix that correspond to each singularity. For robust splitting conditions in particular, an eigenvalue of the scattering matrix is zero. A zero eigenvalue also corresponds to the zero of the determinant, as at least one eigenvalue of $S$ must be $0+i0$ for $det(S)$ to equal $0+i0$. In practice, the conditions for identifying singularities incorporate a tolerance for purposes of visualization.

\begin{figure}[htb]
\hspace*{-0.28cm}
\centering
\includegraphics[width=9.1cm]{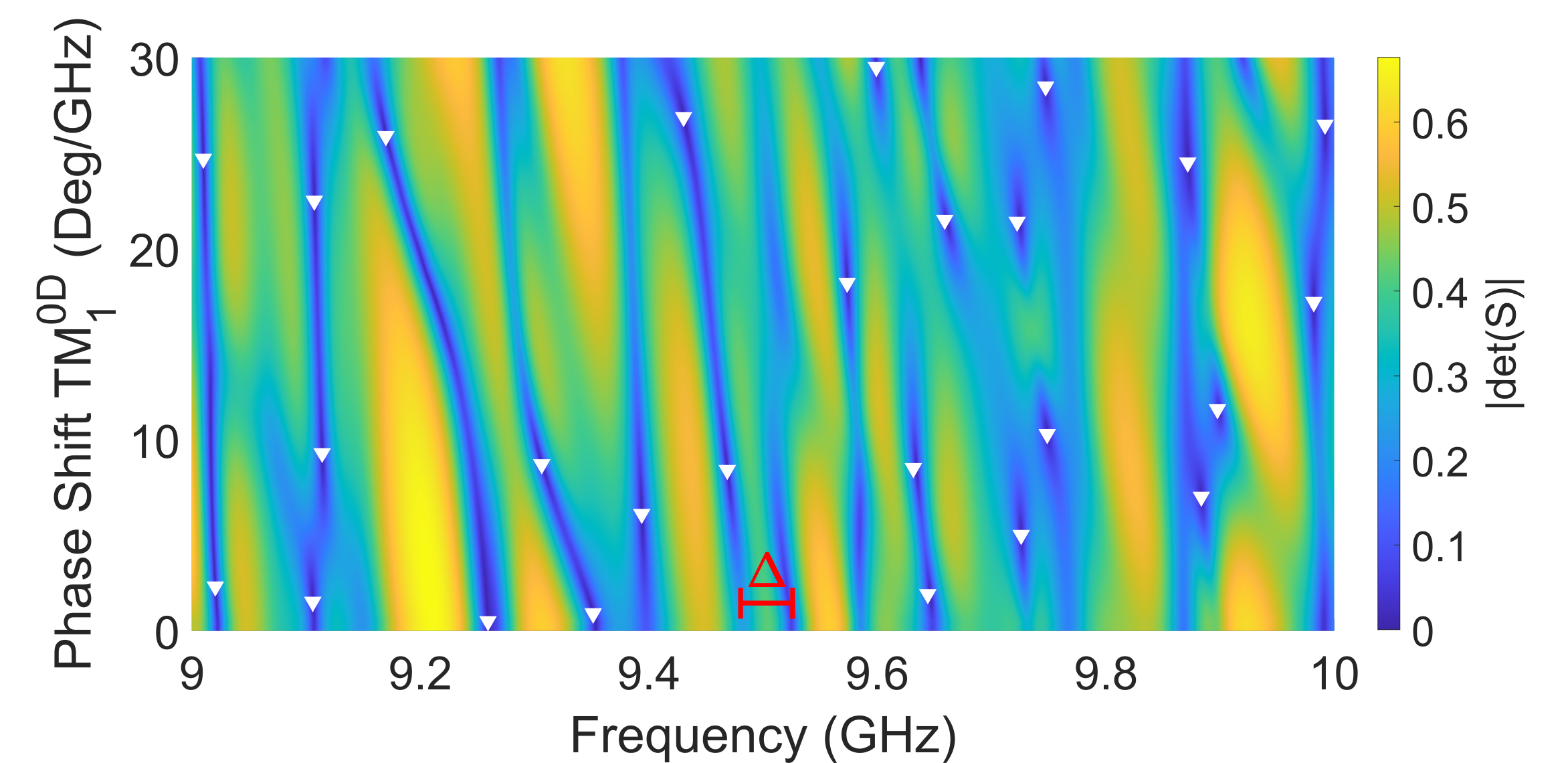}
\caption{Plot of $|det(S)|$ vs frequency and phase shift of $TM_1^{0D}$ in the non-reciprocal tetrahedral microwave graph. The white triangles correspond to points where $det(S)=0+i0$, which enable \rev{robust splitting}. The mean mode spacing ($\Delta=\frac{c}{2L}$) of the graph is approximately 45 MHz, and is marked in red near the bottom of the plot.}
\label{Det_S}
\end{figure}

An abundance of singularities, specifically points of \rev{RS} conditions (corresponding to $det(S)=0+i0$), can be seen in Figure \ref{Det_S} where $|det(S)|$ is plotted vs frequency and phase shift of $TM_1^{0D}$ measured in a non-reciprocal tetrahedral microwave graph. We have identified 26 \rev{RS} conditions marked by white triangles. Numerous \rev{RS} conditions are seen in many types of generic wave-scattering systems with or without reciprocity (see Figs. \ref{Det_S_2D}, \ref{Det_S_3D} in supplementary material \ref{sec.CPA_Surface_Plots}).

Using the 2-port dual-source mode of the VNA, we can directly inject arbitrary monochromatic signals into a system \rev{using one port or both ports simultaneously}. When injecting a signal, the input and output power ratios $\left( \frac{P_1}{P_2} = \frac{\left| V_1 \right|^2}{\left| V_2 \right|^2}\right)$ and phase difference between the ports are systematically varied and measured. From this, we can determine the unique properties of \rev{RS} conditions once we have found their locations in parameter space. Except in the case of exceptional point degeneracies, which is described later and in more detail in supplementary material \ref{sec.Jordan}, any signal sent to, and returned from, an arbitrary two-port scattering system can be written as a linear combination of the two $S$-matrix eigenvectors:

\begin{equation}
\genfrac{(}{)}{0 pt}{}{V_1^{\rm in}}{V_2^{\rm in}} = c_1 \ket{R_{1}} + c_2 \ket{R_{2}} \label{RS_in}
\end{equation}

\begin{equation}
\genfrac{(}{)}{0 pt}{}{V_1^{\rm out}}{V_2^{\rm out}} = S \genfrac{(}{)}{0 pt}{}{V_1^{\rm in}}{V_2^{\rm in}} = c_1 \lambda_S^{(1)} \ket{R_{1}} + c_2 \lambda_S^{(2)} \ket{R_{2}}, \label{S_mat_def}
\end{equation}

where $\left( \genfrac{}{}{0 pt}{}{V_1^{\rm in}}{V_2^{\rm in}} \right)$, $\left( \genfrac{}{}{0 pt}{}{V_1^{\rm out}}{V_2^{\rm out}} \right)$ are the input and output signals respectively, $\ket{R_{1,2}}$ are the right eigenvectors of the sub-unitary scattering matrix $S$, $c_{1,2}$ are arbitrary coefficients, and $\lambda_S^{(1,2)}$ are the corresponding eigenvalues, which are complex in general.

If the system is set to \rev{RS} conditions (\rev{without loss of generality, let} $\lambda_S^{(1)}=0+i0$),

\begin{equation}
\genfrac{(}{)}{0 pt}{}{V_1^{\rm out}}{V_2^{\rm out}} = c_2 \lambda_S^{(2)} \ket{R_{2}}. \label{RS_out}
\end{equation}

Therefore, for any arbitrary monochromatic signals sent into the system, the relative amplitude and phase of the output signals are robust and determined by $\ket{R_{2}}$ (excluding the set of measure zero where $c_1=1$, $c_2=0$, which is the conventional CPA injection \cite{Chen20,Erb24}). The contribution of the eigenvector corresponding to the zero eigenvalue of the input signal is completely absorbed by the system. \rev{The overall amount of signal that is absorbed after injection is dependent on both the proportion of the input signal on $c_1 \ket{R_{1}}$ and the value of $\lambda_S^{(2)}$. For systems with less overall loss, $|\lambda_S^{(2)}|$ will tend to be closer to unity and the output signal will have more power. Additional discussion of signal loss is described in supplementary material \ref{sec.RS_Loss}.} In practice, once the \rev{RS} condition has been found, the robust splitting value that the output signal obeys can be indirectly determined from the relative amplitude and phase of the two components of the scattering matrix eigenvector corresponding to the non-zero eigenvalue (see supplementary material \ref{sec.CPA_Form} for the analytic form of the eigenvectors at \rev{RS} conditions). \rev{This eigenvector depends on the scattering matrix elements, so in general at different \rev{RS} conditions, the output power ratio and phase difference will also be different.}

To directly determine the robust splitting ratio of \rev{RS} conditions, we inject signals over a large range of channel amplitudes and relative phases. In Figure \ref{Robust_CPA}, we experimentally demonstrate that the power ratio and phase difference of the output signals at an \rev{RS} condition are robust to 100 dB of relative power and $2\pi$ phase change of the input signal. \rev{Note, for input power ratios $|\text{P}_1^{\:\rm in}/ \text{P}_2^{\:\rm in}| \ge 40$ dB, this is effectively a single port excitation.} In this figure, for the specific \rev{RS} condition measured, the amplitude and phase of the output signals are fixed at approximately 3.85 and -128.8 degrees respectively, independent of the value of the input signals.

\begin{figure}[htb]
\centering
\includegraphics[width=8.7cm]{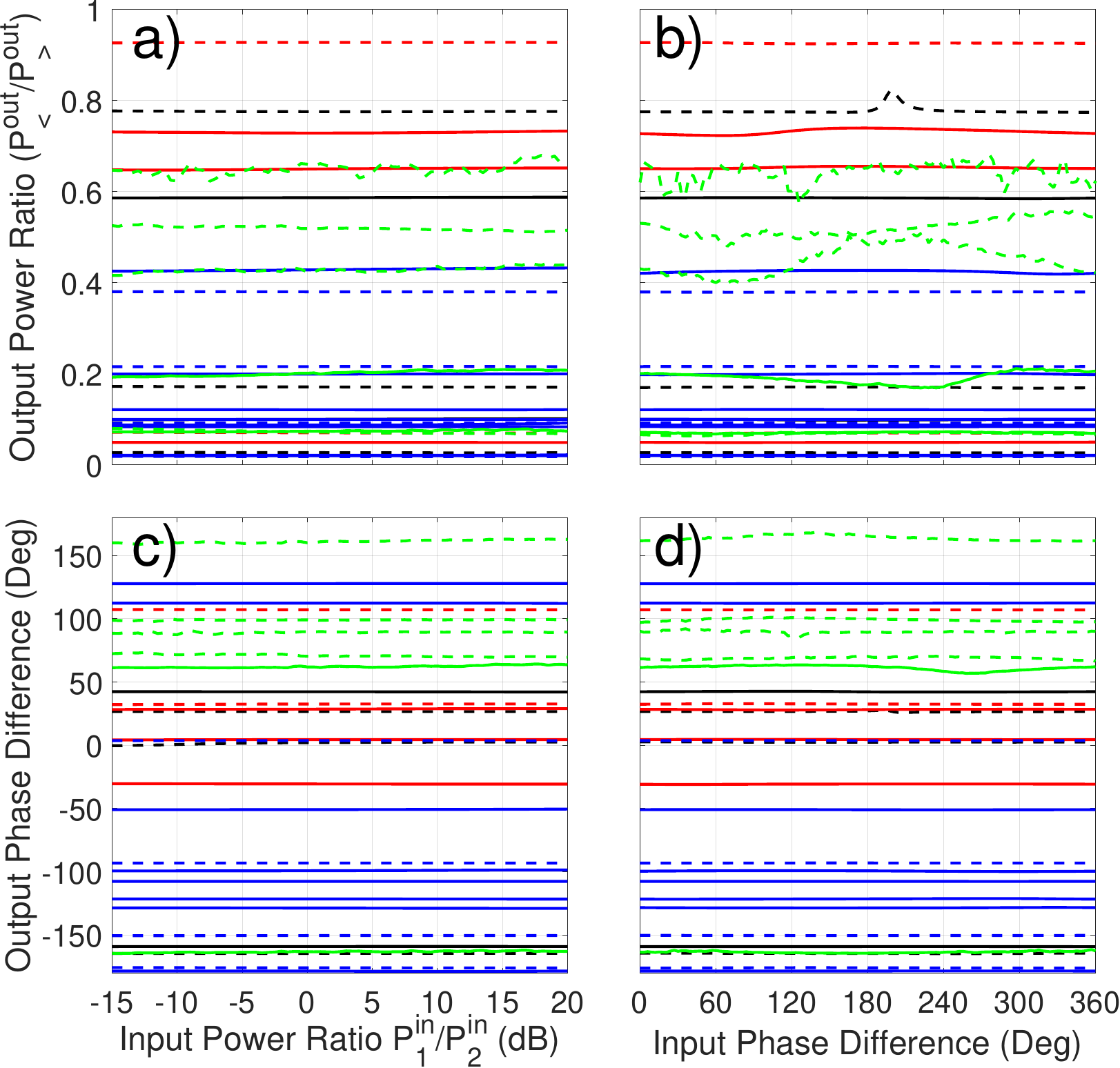}
\caption{Experimental demonstration of the robust splitting of 28 unique \rev{robust splitting} conditions from four different experimental systems. The black curves were measured in a 1D reciprocal tetrahedral graph, red curves in a 1D non-reciprocal tetrahedral graph, blue curves in a 2D quarter bow-tie billiard, and green curves in a chaotic 3D microwave cavity. The dashed lines correspond to output power ratios where $P_1^{\:\rm out} > P_2^{\:\rm out}$ and solid lines correspond to $P_1^{\:\rm out} < P_2^{\:\rm out}$. (a) Output power ratio vs input power ratio. (b) Output power ratio vs input phase difference. (c) Output phase difference vs input power ratio. (d) Output phase difference vs input phase difference.}
\label{Multiple_CPA}
\end{figure}

To show that different \rev{RS} conditions have different splitting ratios, we found 28 unique \rev{RS} conditions in all four of the physical systems described above, and show the results in Figure \ref{Multiple_CPA}. In this figure, we show the output power ratio and phase difference vs the input power ratio and input phase difference for all 28 \rev{RS} conditions measured. The color of the curve corresponds to the specific system measured (black: reciprocal graph, red: non-reciprocal graph, blue: two-dimensional billiard, green: three-dimensional cavity). The solid curves correspond to measurements where the power at port 1 was less than the power at port 2, and vice versa for the dashed curves. We see that nearly all curves are flat, proving the robustness of the output signals at \rev{RS} conditions. The green curves from the three-dimensional cavity are the least flat because during the time it takes to find the \rev{RS} conditions and inject the input signals, the environment of the system can change, causing the system to drift away from the \rev{RS} conditions. See Figure \ref{Arbitrary_Injection} in supplementary material \ref{sec.Arbitrary_Inj} to see signal injection in systems not set to \rev{RS} conditions. In contrast with the robustness of the output signals at \rev{RS} conditions, output signals away from \rev{RS} conditions vary significantly as the input signals change.

\begin{figure}[htb]
\centering
\includegraphics[width=8.7cm]{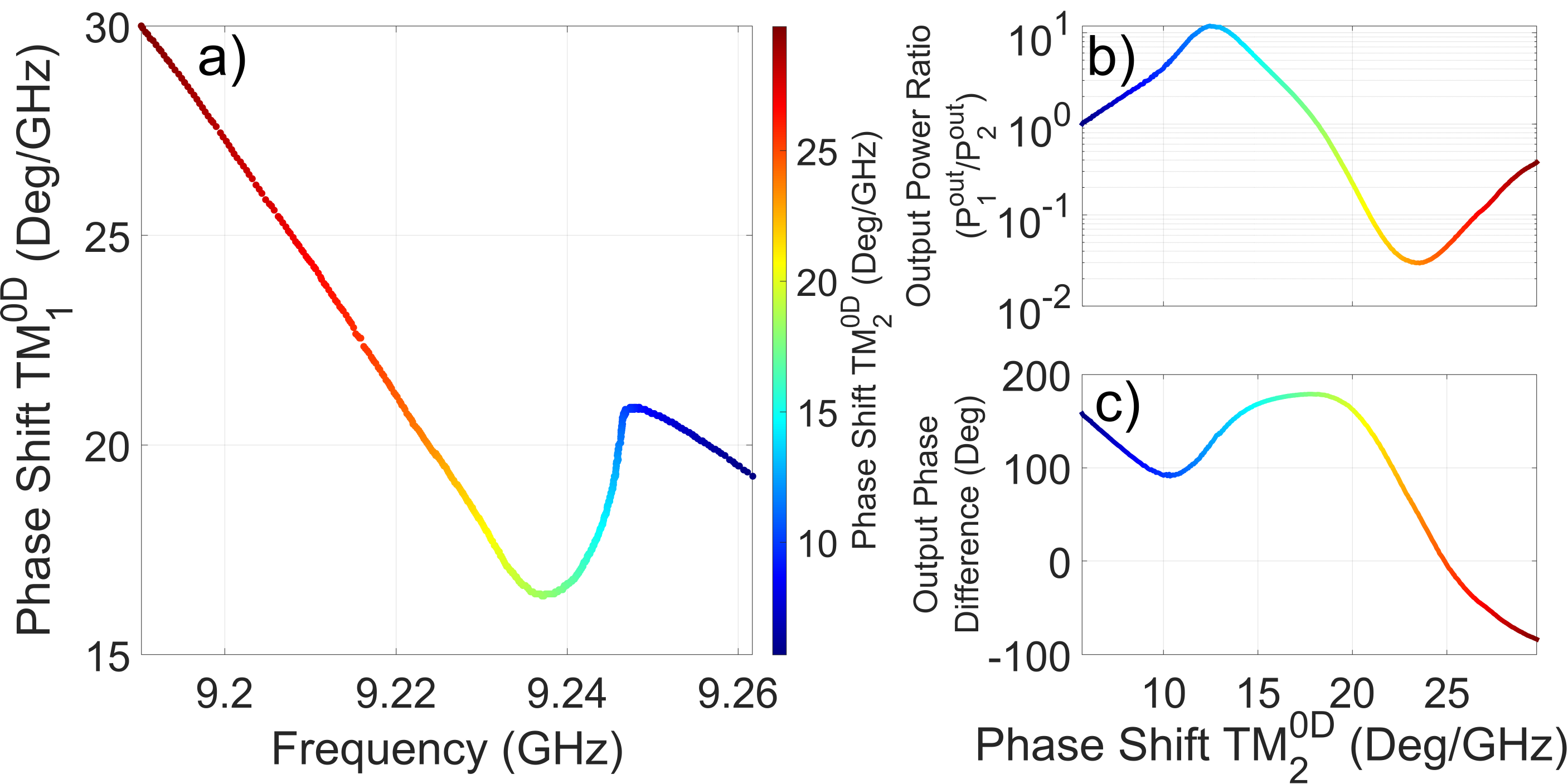}
\caption{Experimental demonstration of the continuous tunability of \rev{robust splitting} conditions in the reciprocal tetrahedral microwave graph. The color of each point corresponds to the value of the varying third parameter of the system ($TM_2^{0D}$) the \rev{RS} condition was measured at. (a) Location of an \rev{RS} condition in the two-dimensional parameter space of frequency and phase shift of $TM_1^{0D}$ as the phase shift of $TM_2^{0D}$ varies. (b) Output power ratio vs phase shift of $TM_2^{0D}$. (c) Output phase difference vs phase shift of $TM_2^{0D}$.}
\label{Tunable_CPA}
\end{figure}

If we slowly vary the value of a third tunable parameter while measuring the scattering matrix over the same two-dimensional parameter space, such as the one shown in Fig. \ref{Det_S}, we can continuously manipulate the location of the singularities of the scattering matrix \cite{Erb25} (see supplementary material \ref{sec.Movie} for a video showing the dynamics of \rev{RS} conditions). Within the two-dimensional parameter space, as the third parameter varies, the scattering singularities are topologically stable except in certain instances when they are created or annihilated in pairs. For \rev{RS} conditions in particular, as their locations move in parameter space, their robust splitting values also vary. This is shown in Figure \ref{Tunable_CPA}, where the left plot shows the location of an \rev{RS} condition moving in the frequency and phase shift of $TM_1^{0D}$ parameter space as the phase shift of $TM_2^{0D}$ is varied. The right plots show how the output power ratio and phase difference vary along the trajectory of the evolving \rev{RS} condition. The colors show the continuous evolution of the third parameter, phase shift of $TM_2^{0D}$. For this \rev{RS} condition, the output power ratio varies by multiple orders of magnitude and the output phase difference varies over 260 degrees. Similar examples obtained in the two-dimensional quarter bow-tie billiard and the three-dimensional microwave cavity are shown in Figs. \ref{Tunable_CPA_2D}, \ref{Tunable_CPA_3D}. \rev{Due to the complexity of our experimental systems, any perturbation to the system will affect the scattering properties, which also affects the location of the \rev{RS} conditions as well as their output power ratio and phase difference.} To view the stability of \rev{RS} conditions over time in a one- and two-dimensional system see Figs. \ref{Fixed_CPA_Graph}, \ref{Fixed_CPA_Billiard} in supplementary material \ref{sec.Stability}. 

\begin{figure}[htb]
\centering
\includegraphics[width=8.7cm]{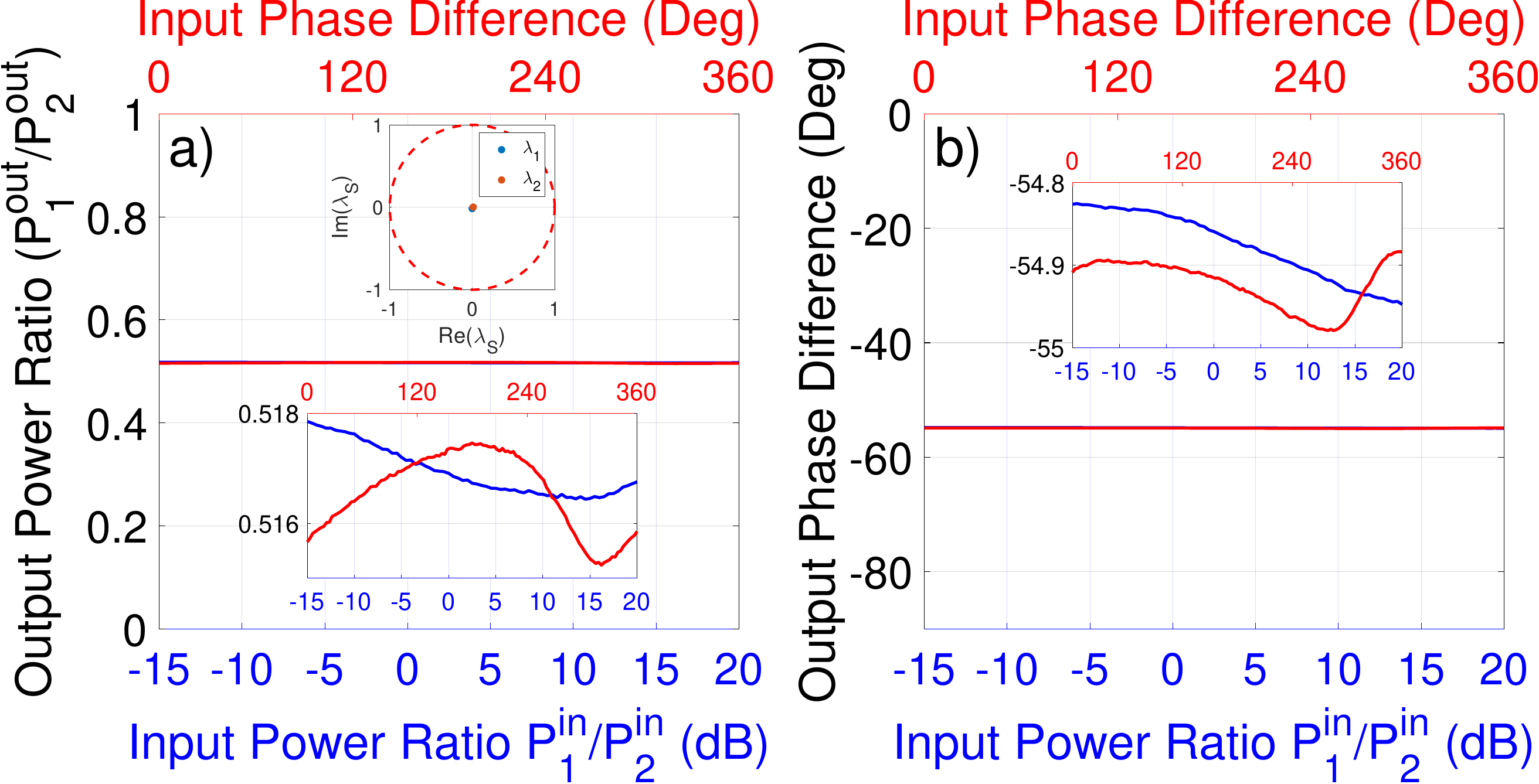}
\caption{Experimental demonstration of the robust splitting of \rev{RS}+EPD in a non-reciprocal tetrahedral microwave graph. (a) Output power ratio vs input power ratio (lower axis, blue) and input phase difference (upper axis, red). The blue curves correspond to the input signals power ratio being swept, and the red curves correspond to the input signals relative phase being swept. Top inset shows the eigenvalues of the scattering matrix in the complex plane. Bottom inset shows a vertical zoom-in to illustrate the details of the curves. (b) Output phase difference vs input power ratio (lower axis, blue) and input phase difference (upper axis, red). The blue and red curves have the same interpretation as in (a). Inset shows a vertical zoom-in to magnify the details of the curves.}
\label{CPA_EP}
\end{figure}

We have established that utilizing a single tunable parameter and frequency we can find an abundance of scattering singularities. However, within each of the physical systems, there are multiple phase shifters or metasurfaces which can be used to perturb the system. Using these multiple tunable parameters, we can do more advanced manipulation of the scattering singularities, such as combining them at a single location in a two-dimensional parameter space. A particularly interesting combination case is that of an exceptional point degeneracy (EPD) and a \rev{robust splitting} condition. At an exceptional point degeneracy, both the scattering matrix eigenvalues and associated eigenvectors become degenerate \cite{Dembowski01,Berry2004,Lee08,Wiersig14,Alu19,Özdemir19}. In prior work, the CPA and EPD combination has been shown in the spectrum of the Hamiltonian \cite{Stone19,Yang21,Farhi22} and the \rev{RS} condition and EPD combination has been shown in the scattering matrix of a reciprocal scattering system \cite{Erb25}. In Figure \ref{CPA_EP}, we demonstrate an \rev{RS}+EPD combination in a non-reciprocal system for the first time. In this figure, we show that both scattering eigenvalues are near $\lambda_S = 0+i0$, and that the output signal power ratio and phase difference are robust to changes in the input signal power ratio and phase difference. The results show small deviations from ideal flatness because the scattering eigenvalues $|\lambda_S^{(1)}|+|\lambda_S^{(2)}|=0.029 \neq 0$. As this was measured in a non-reciprocal system, the combined case of \rev{RS} conditions and EPD is no longer a 50:50 power splitter, as predicted for reciprocal systems in Ref. \citenum{Erb25}. The output power ratio and phase difference can take any value in non-reciprocal systems, as can be seen from the analytic formula for the degenerate eigenvector at \rev{RS} and EPD conditions: $\ket{R_{RS+EPD}} \propto
\begin{pmatrix} \frac{-S_{22}}{S_{21}}\\ 1 \end{pmatrix}$ (see supplementary material \ref{sec.Jordan} for additional details on the RS and EPD combination). \rev{For each instance of an \rev{RS}+EPD combination in a non-reciprocal system, the value of $\frac{-S_{22}}{S_{21}}$ (magnitude and phase) fixes the output power ratio and phase difference, but different instances of \rev{RS}+EPD combinations will have different values of $\frac{-S_{22}}{S_{21}}$. In addition, since the \rev{RS} conditions and EPD's are independent singularities, in general any perturbation to the system will split the combination apart.}

All of the above results are generic to all complex wave-scattering systems. To confirm this, we \rev{utilize} a Random Matrix Theory model (see supplementary material \ref{sec.RMT}, \ref{sec.RMT_Results}), \rev{which have been well established in the literature to describe the universal scattering properties of generic complex resonant systems \cite{Couchman1992,Fyodorov1997bTRI,Beenakker1997,Mendez2003,Fyodorov2004,Kuhl2005,FSav11,Kuhl2013,Li2017,Grabsch2018,Hougne2020}.} In Figure \ref{Det_S_RMT}, similar to Figure \ref{Det_S}, the $|det(S)|$ is plotted vs $\omega$ and parameter $x$, where $\omega$ is the frequency and parameter $x$ is a surrogate for the tunable metasurface. There are 21 \rev{RS} conditions marked by white triangles. In Figure \ref{Tunable_CPA_RMT}, similar to Figure \ref{Tunable_CPA}, we see that the location of an \rev{RS} condition can be strongly manipulated in a two-dimensional parameter space by means of a third parameter (\rev{in this case parameter} $y$, \rev{which acts similarly to $x$}), and the output power ratio and phase difference vary significantly.


In conventional RF and microwave settings, there are several passive devices that act as a wave splitter. These devices include power splitters, directional couplers, hybrid couplers, etc \cite{Pozar11}. Most commonly, these devices take the form of a three- or four-port network, with the input ports being distinct from the output ports. Examples of power splitters are T-junctions and Wilkinson power dividers \cite{Wilkinson60}. Generally, power splitters take an input signal and split it into two output signals with equal amplitude and phase, although it's possible for output signals to be split into $N$ channels with an unequal power split. Directional couplers can be designed for an arbitrary (but fixed) output power split, while hybrid couplers generally have an equal power split \cite{Pozar11}. The $90^\circ$ and $180^\circ$ hybrid couplers take an input signal and split it into two equal amplitude signals with a $90^\circ$ phase difference or either a $180^\circ$ or $0^\circ$ phase difference respectively.

In general, conventional wave splitters are networks (graphs) that take a single input signal and output $N\ge2$ signals at some fixed amplitude and phase ratios, where the input and output ports are distinct. These splitters can be designed to work over a broad frequency band, but the output amplitudes and phases cannot be tuned. In contrast, we have shown that the \rev{RS} phenomenon can convert any generic two-port complex scattering system with adjustable parameters into a \textit{tunable} robust splitter. \rev{One should note that the output signals suffer a degree of loss dependent on the value of the non-zero eigenvalue and the details of the input signal (see supplementary material \ref{sec.RS_Loss}).} Additionally, this phenomenon only works at a single frequency at a time, but due to the abundance and manipulability of \rev{RS} conditions, a tunable system can exhibit robust splitting at arbitrary frequencies. Another important distinction is that for \rev{RS} conditions, the input ports and output ports are the same. The input signal can be injected into one or both ports, but the output signal will have a robust splitting value between both ports. Due to the directed nature of conventional splitters, there is usually unwanted reflection at the input port. However in the \rev{RS} case, the input and output ports are identical, therefore the issue of unwanted reflection is irrelevant. 

The results discussed in this work were obtained using a $2\times 2$ scattering matrix, but can be generalized to higher dimensions. The conditions for \rev{$det(S)=0+i0$} are still abundant, topologically protected, and can be manipulated in higher dimensions as $det(S)$ remains a simple complex scalar function for any number of ports. Unfortunately, the robust splitting phenomenon only generically occurs when measuring a $2\times 2$ scattering matrix (see supplementary material \ref{sec.Jordan_Higher} for details on signal injection in scattering systems \rev{with many ports}). However, systems with multiple ports can still achieve robust splitting between any two ports by only injecting and receiving signals on those two ports, while also at \rev{RS} conditions.

Instead of the conventional use of CPA to absorb all incident energy, we have discovered a way to use \rev{RS} conditions to create robust splitters \rev{(albeit with a variable and sometimes substantial overall loss)}, and with multiple tunable parameters we can continuously change the relative amplitude and phase of the robust output signal. Using enough tunable parameters, a degree of control can be created over the output splitting values, \rev{including independent control of the output power ratio and phase difference}. With the ability to generate and manipulate numerous scattering singularities, this provides an alternative to special design/engineering considerations to \rev{establish RS conditions} at particular frequencies, since the tunable parameters take care of that problem. The topological protection of the \rev{RS} conditions (as long as they are not annihilated with a separate \rev{RS} condition), and their ease of manipulation, make applications of these singularities more feasible. Additionally, using the RMT model we have shown that these phenomena are general to all complex wave-scattering systems.


\begin{acknowledgments}
This work was partially supported by NSF/RINGS under grant No. ECCS-2148318, ONR under grant N000142312507, DARPA WARDEN under grant HR00112120021, Department of Energy under grant DE-SC0024223, and US-Israel Binational Science Foundation under grant BSF2022158. 
\end{acknowledgments}

\section*{Data Availability Statement}

\rev{The data that support the findings of this study are openly available in the Digital Repository at the University of Maryland at https://doi.org/10.13016/lxva-rkcs.}

\clearpage
\newpage

\section*{Supplementary Material}

In this supplementary material, section \ref{sec.Exp} describes the experimental systems \rev{and procedures} used in this work, section \ref{sec.CPA_Surface_Plots} demonstrates \rev{robust splitting (RS)} phenomena in the experimental systems not shown in the main text, section \ref{sec.Jordan} describes exceptional point degeneracy (EPD) phenomena and how it affects signal injection, \rev{section \ref{sec.RS_Loss} discusses the overall loss associated with signals injected at RS conditions}, section \ref{sec.CPA_Form} gives the general analytic form of the eigenvectors and their simplified forms at \rev{RS} conditions, section \ref{sec.Arbitrary_Inj} shows signal injection into systems not at \rev{RS} conditions, section \ref{sec.Movie} contains a movie of the dynamics of \rev{RS} conditions, section \ref{sec.Stability} discusses the stability over time of systems set to \rev{RS} conditions, section \ref{sec.RMT} details the Random Matrix Theory (RMT) model of complex scattering systems used in this work, section \ref{sec.RMT_Results} shows results of the RMT model that have the same features as the experimental results, and section \ref{sec.Jordan_Higher} generalizes signal injection and EPD's for higher dimensional scattering matrices.

\setcounter{figure}{0}
\setcounter{equation}{0}
\setcounter{section}{0}

\renewcommand{\figurename}{Fig.}
\renewcommand{\thefigure}{S\arabic{figure}}
\renewcommand{\theequation}{S\arabic{equation}} 
\renewcommand{\theHfigure}{S\arabic{figure}}
\renewcommand{\theHequation}{S\arabic{equation}}

\rev{\section{Measurement Procedure and Experimental Scattering Systems}\label{sec.Exp}}

\rev{To find RS conditions experimentally, we parametrically measure the scattering matrix $S$ vs tunable parameters, usually frequency and metasurface applied bias voltage or phase shift of a phase shifter. Next we analyze the scattering matrices of the entire parameter space measured and find the particular parameter settings which result in a value of $|det(S)|$ smaller than some threshold value (usually 0.002), corresponding to a candidate RS condition. Choosing a particular candidate RS condition, if necessary, we perform fine sweeps of each tunable parameter until the value of $|det(S)|$ is minimized. The smaller the value of $|det(S)|$ is, the more robust the output splitting ratios will be. In the next step, keeping the system fixed at the optimal parameters, we switch the microwave VNA to the 2-port dual-source mode, so that we can inject arbitrary signals into the system using one port or both ports simultaneously, and measure the output signals from the system on both ports. Then we inject the arbitrary signals at the RS condition over a wide range of relative amplitudes and phases to demonstrate that the output signal power ratio and relative phase between both ports are fixed.}

\subsection{One-dimensional Graphs}
Networks of coaxial cables (graphs) consist of bonds and nodes, where the one-dimensional (1D) Helmholtz equation describes wave propagation along the bonds \cite{Hul04,Lawniczak10,Shaibe25,Erb25}. We utilize a quasi-one-dimensional tetrahedral microwave graph with tunable phase shifters along four of the bonds (see inset (i) of Fig. \ref{Robust_CPA}(a)). Each Narda-MITEQ P1507D-SM24 phase shifter can change the effective length of a bond by approximately 5 cm. The reciprocal (non-reciprocal) graph has a total electrical length of approximately 3.2 m (3.3 m). For the non-reciprocal graph, an internal T-junction node is replaced with a Narda-MITEQ Model 4925 Circulator.

\subsection{Two-dimensional Billiard}
The quasi-two-dimensional system utilized in this work is a brass quarter bow-tie billiard with two coupling ports on the lid \cite{Gokir98,Wu98,Chung00} (see inset (ii) of Fig. \ref{Robust_CPA}(a)). When the billiard is excited below frequencies of approximately 19 GHz, only a single propagating mode is supported, with the electric field polarized in the vertical (thin) direction \cite{So95,Stock99}. The billiard has an area of 0.115 $m^2$ and a height of 7.9 mm. Inside the billiard, there are three tunable metasurfaces fabricated by Johns Hopkins University Applied Physics Laboratory \cite{PhysRevApplied.20.014004} and designed to vary reflection amplitude (phase) between 11-18 GHz (14-16 GHz). The metasurfaces are composed of 18 mushroom-shaped resonant elements in a linear array, where each element is sub-wavelength in size and loaded with varactor diodes. Each metasurface is 7.9 mm high, 185 mm long, and 1.8 mm thick. On a given metasurface, all diodes are tuned simultaneously by applying a global DC voltage bias. Generally, both the reflection magnitude and phase of the metasurface change as the voltage is varied. Each metasurface covers approximately 12\% of the perimeter of the billiard. 

\subsection{Three-dimensional Cavity}
The three-dimensional system utilized in this work is a nearly-cubic cavity connected to the outside world through two ports (see inset (iii) of Fig. \ref{Robust_CPA}(b)). The cavity has a volume of $\sim$0.76 $m^3$ and side lengths of approximately 0.92 m \cite{Frazier20,Frazier22}. There are two two-dimensional metasurfaces inside the cavity and various irregularly-shaped scatterers to increase the complexity of the system. The two-dimensional metasurfaces were fabricated by Johns Hopkins University Applied Physics Laboratory \cite{PhysRevApplied.20.014004} and were designed to vary reflection amplitude (phase) between 2-3.6 GHz (3-3.5 GHz). The metasurfaces are composed of $14\times 14$ mushroom-shaped resonant elements in a square array, where each element is sub-wavelength in size and loaded with four varactor diodes. Each metasurface is 26 cm by 26 cm, and cover approximately 1.3\% of the interior surface area of the cavity. On a given metasurface, all diodes are tuned simultaneously by applying a global DC voltage bias. Generally, both the reflection magnitude and phase of the metasurface change as the voltage is varied.

The one- and two-dimensional tunable metasurfaces are operated in the low-power linear-response limit. Under these conditions, the scattering properties of the system, including the varactor diodes, are fully captured by the linear scattering matrix.

\begin{figure}[htb]
\hspace*{-0.28cm}
\centering
\includegraphics[width=9.1cm]{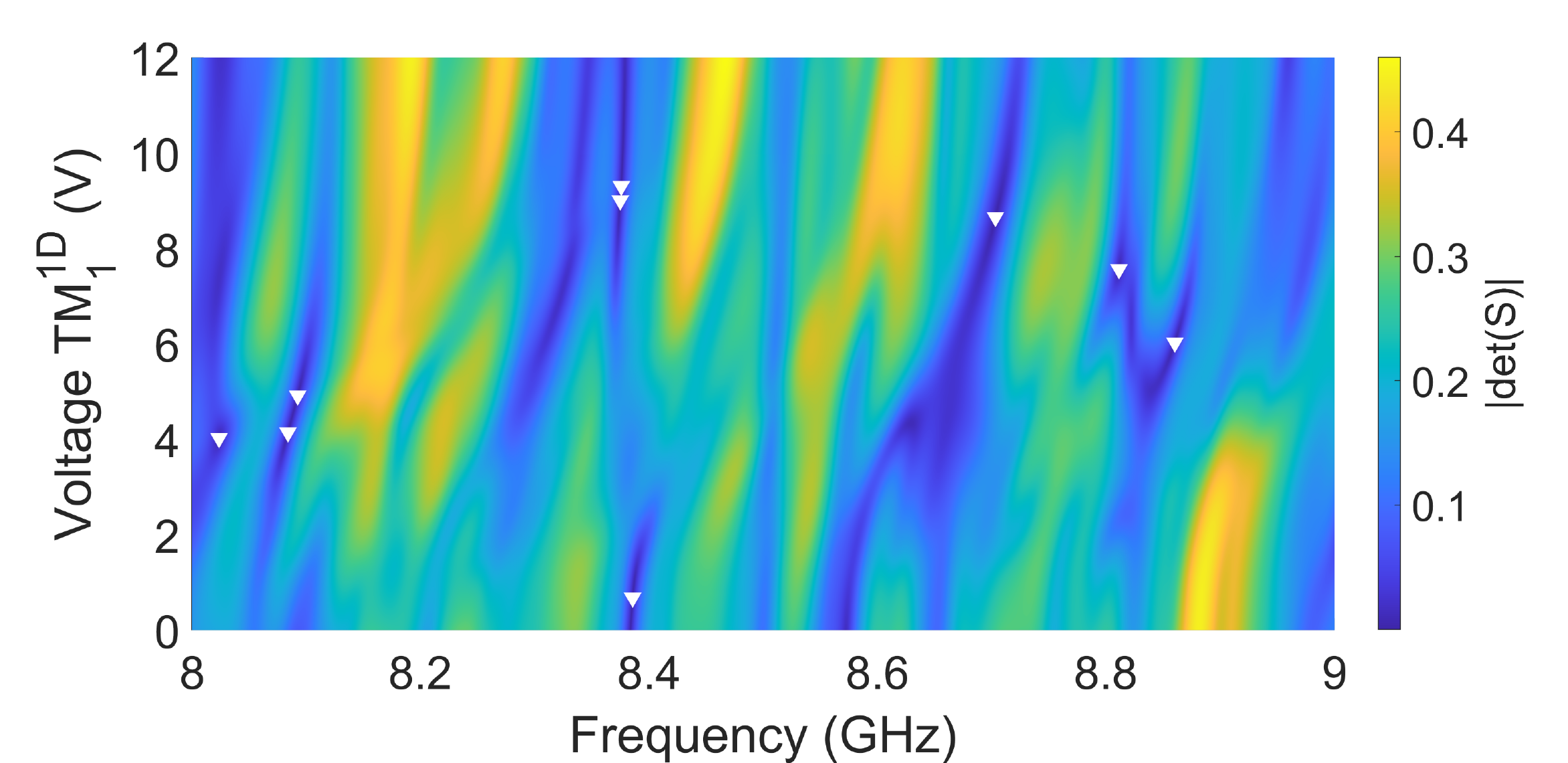}
\caption{Plot of $|det(S)|$ vs frequency and $TM_1^{1D}$ applied bias voltage in the quarter bow-tie two-dimensional microwave billiard. The white triangles correspond to points where $det(S)=0+i0$, which enable \rev{robust splitting}.}
\label{Det_S_2D}
\end{figure}

\begin{figure}[htb]
\hspace*{-0.23cm}
\centering
\includegraphics[width=8.9cm]{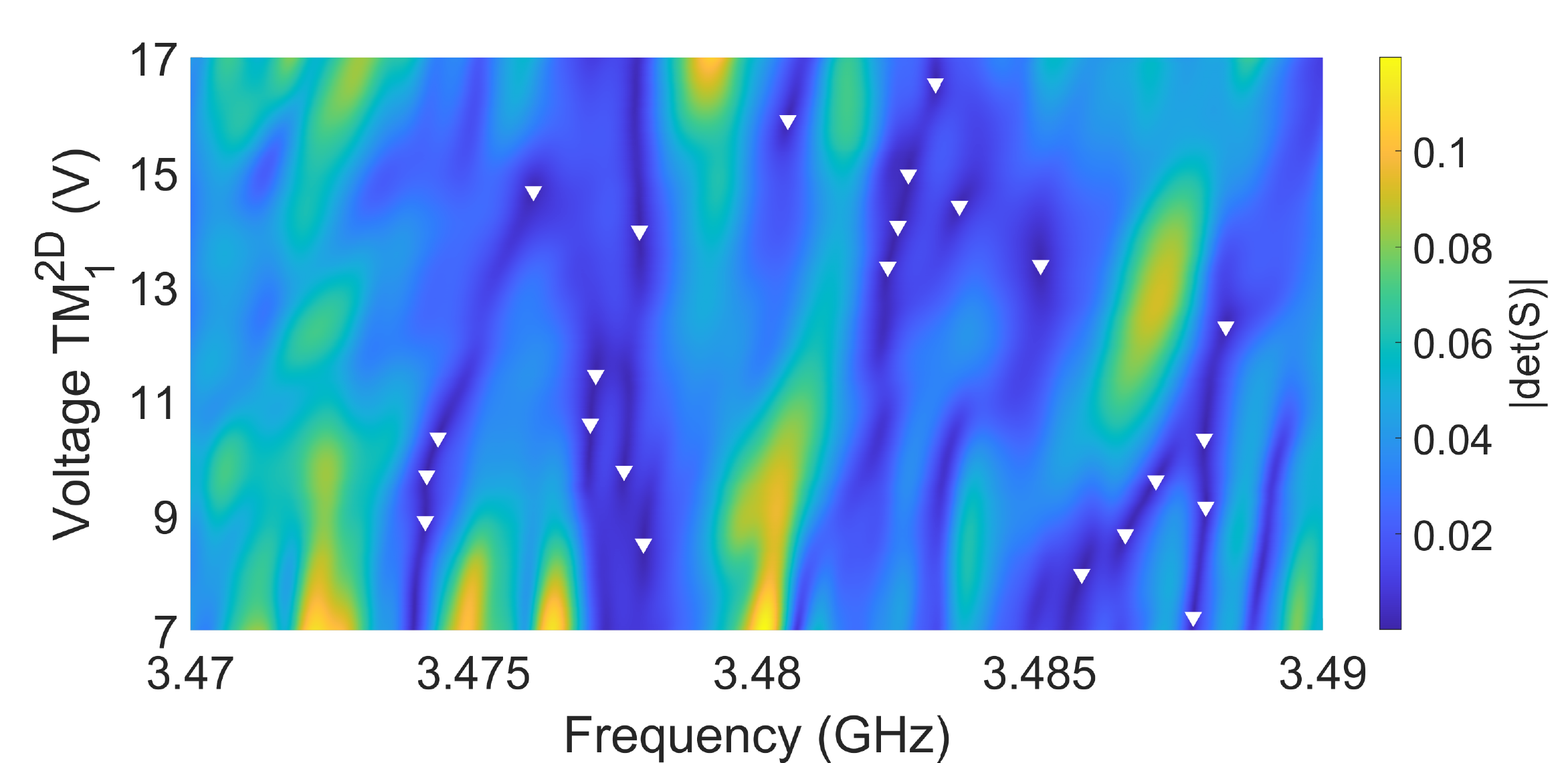}
\caption{Plot of $|det(S)|$ vs frequency and $TM_1^{2D}$ applied bias voltage in the three-dimensional microwave cavity. The white triangles correspond to points where $det(S)=0+i0$, which enable \rev{robust splitting}.}
\label{Det_S_3D}
\end{figure}

\begin{figure}[htb]
\centering
\includegraphics[width=8.7cm]{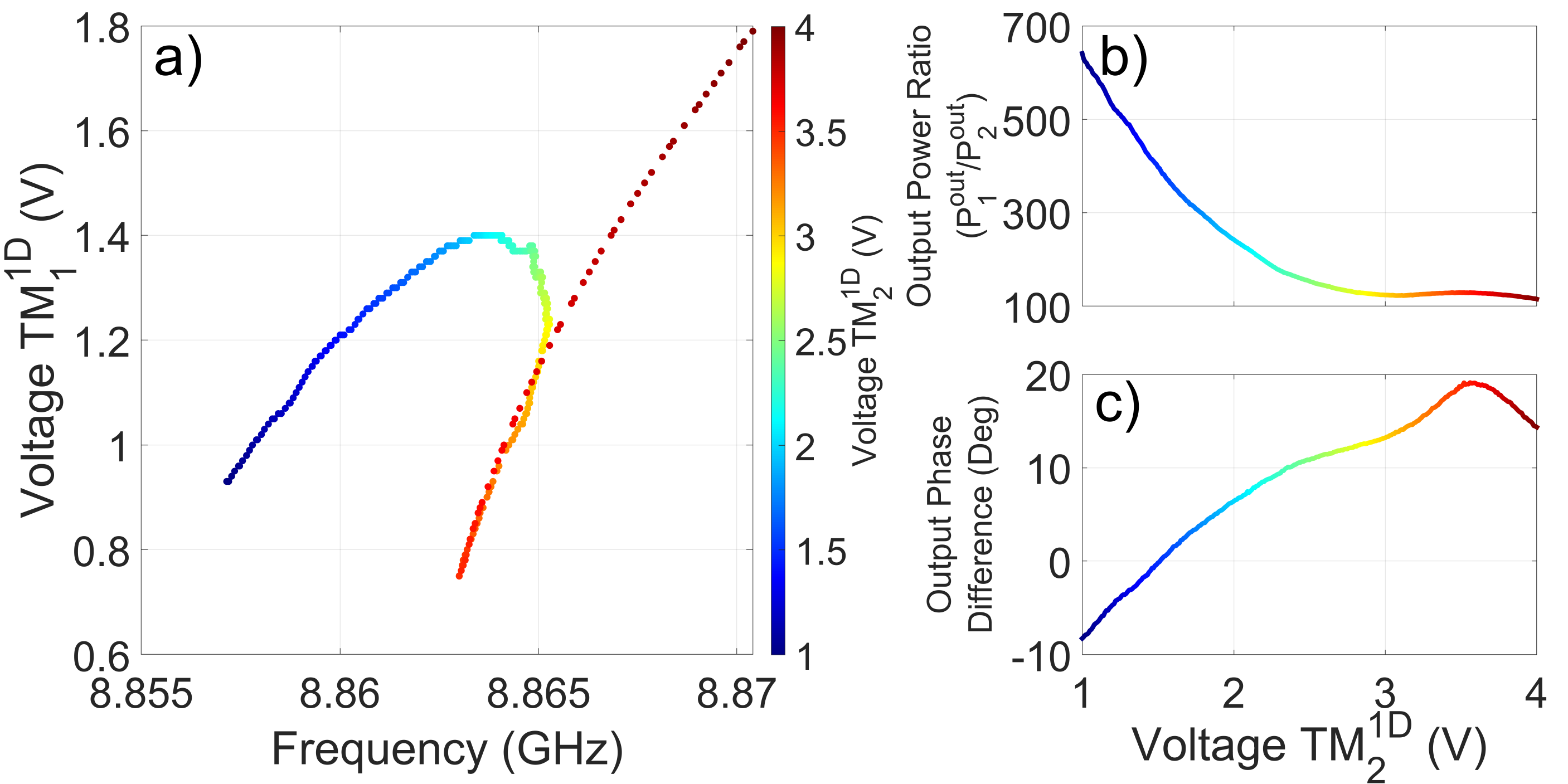}
\caption{Experimental demonstration of the tunability of \rev{robust splitting} conditions in the quarter bow-tie two-dimensional microwave billiard. The color of each point corresponds to the value of the varying third parameter of the system ($TM_2^{1D}$) the \rev{RS} condition was measured at. (a) Location of an \rev{RS} condition in the two-dimensional parameter space of frequency and $TM_1^{1D}$ applied bias voltage as the applied bias voltage of $TM_2^{1D}$ varies. (b) Output power ratio vs $TM_2^{1D}$ applied bias voltage. (c) Output phase difference vs $TM_2^{1D}$ applied bias voltage.}
\label{Tunable_CPA_2D}
\end{figure}

\begin{figure}[htb]
\centering
\includegraphics[width=8.7cm]{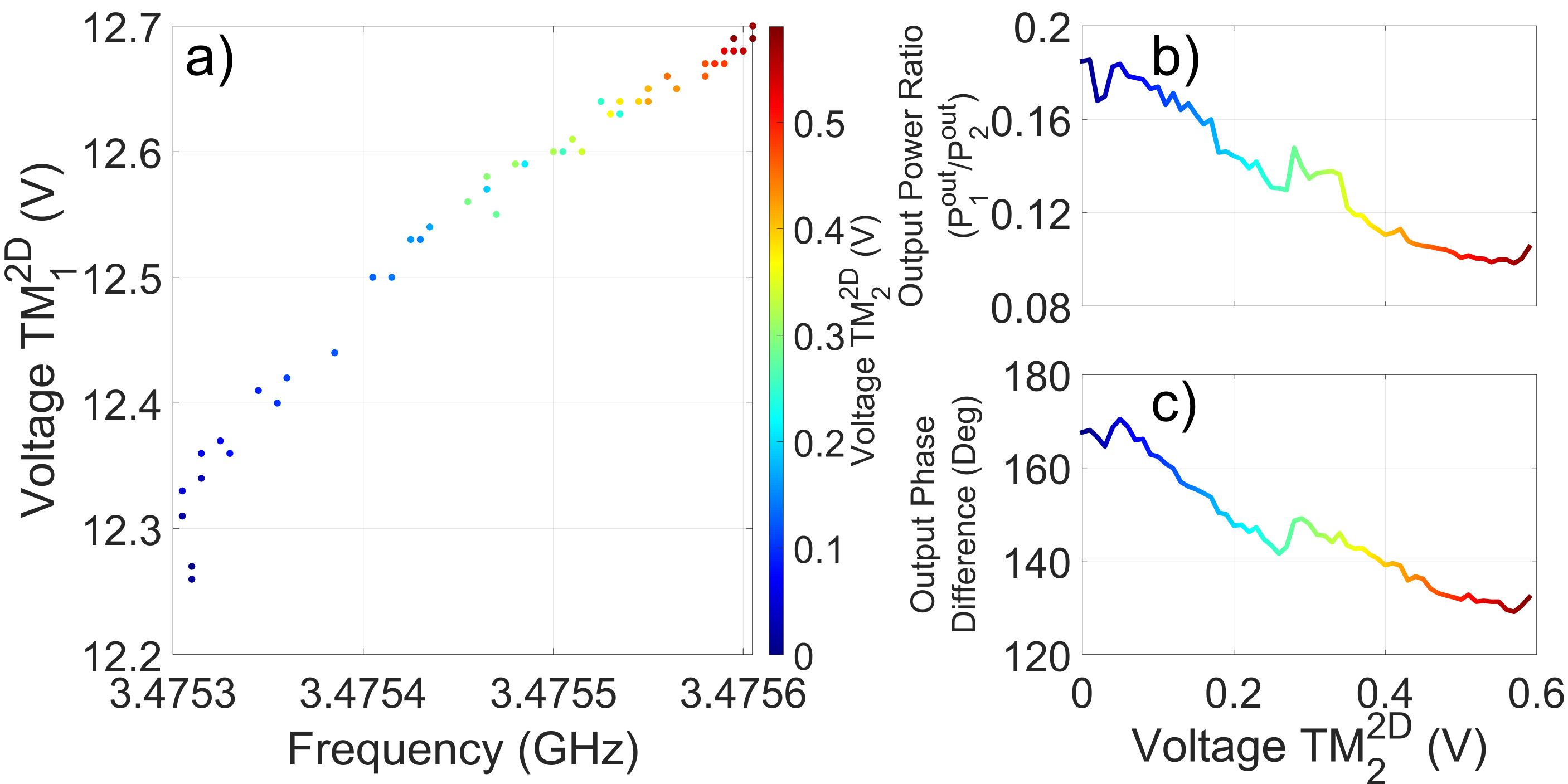}
\caption{Experimental demonstration of the tunability of \rev{robust splitting} conditions in the three-dimensional microwave cavity. The color of each point corresponds to the value of the varying third parameter of the system ($TM_2^{2D}$) the \rev{RS} condition was measured at. (a) Location of an \rev{RS} condition in the two-dimensional parameter space of frequency and $TM_1^{2D}$ applied bias voltage as the applied bias voltage of $TM_2^{2D}$ varies. (b) Output power ratio vs $TM_2^{2D}$ applied bias voltage. (c) Output phase difference vs $TM_2^{2D}$ applied bias voltage.}
\label{Tunable_CPA_3D}
\end{figure}

\rev{\section{Robust Splitting Conditions in Other Physical Systems}\label{sec.CPA_Surface_Plots}}
All physical systems measured in this work have similar dynamics and structure of $|det(S)|$ in two-dimensional parameter spaces, where there are numerous \rev{RS} conditions which can be strongly manipulated as the system is tuned by a third parameter. In the main text, we showed an example of $|det(S)|$ vs two tunable parameters and an example of an \rev{RS} condition being manipulated in parameter space for tetrahedral microwave graphs (Figs. \ref{Det_S}, \ref{Tunable_CPA}). Here we show the same results, but for the quarter bow-tie two-dimensional billiard (Figs. \ref{Det_S_2D}, \ref{Tunable_CPA_2D}) and the three-dimensional microwave cavity (Figs. \ref{Det_S_3D}, \ref{Tunable_CPA_3D}).

\section{Signal Injection at Exceptional Point Degeneracy and Robust Splitting Conditions} \label{sec.Jordan}
At an exceptional point degeneracy, the scattering matrix eigenvalues and eigenvectors coalesce resulting in a collapse of the eigenbasis. To complete the eigenbasis, a Jordan vector $\ket{J_{EPD}}$ is introduced utilizing the Jordan chain scheme, leading to this expression for the Jordan vector \cite{Bronson69,Seyranian03}: 

\begin{equation}
S \ket{J_{EPD}} = \lambda_{S}^{EPD} \ket{J_{EPD}} + \ket{R_{EPD}}, 
\label{JV}
\end{equation}

where $\lambda_{S}^{EPD} = \frac{S_{11}+S_{22}}{2}$, $\ket{R_{EPD}} = \begin{pmatrix} \frac{S_{11}-S_{22}}{2S_{21}} \\ 1 \end{pmatrix}$ are the degenerate eigenvalues and eigenvectors respectively. 

When injecting arbitrary signals into a system at EPD conditions, the input and output signals become: 

\begin{equation}
\genfrac{(}{)}{0 pt}{}{V_1^{\rm in}}{V_2^{\rm in}} = c_1 \ket{R_{EPD}} + c_2 \ket{J_{EPD}}
\end{equation}

\begin{equation}
\genfrac{(}{)}{0 pt}{}{V_1^{\rm out}}{V_2^{\rm out}} = \lambda_S^{EPD} \left[c_1 \ket{R_{EPD}} + c_2 \ket{J_{EPD}}\right] + c_2 \ket{R_{EPD}}. \label{eqn:EPD_output}
\end{equation}

From Eq. (\ref{eqn:EPD_output}), it is clear that if the degenerate eigenvalue $\lambda_S^{EPD}$ is zero (which corresponds to a combined EPD and \rev{RS} condition), then the outgoing wave will take the simple form $\left( \genfrac{}{}{0 pt}{}{V_1^{\rm out}}{V_2^{\rm out}} \right) = c_2 \ket{R_{EPD}}$. To determine the form that $\ket{J_{EPD}}$ takes at \rev{RS} conditions, we rewrite Equation (\ref{JV}) in a different form: 

\begin{equation}
(S - \lambda_S^{RS+EPD} I) \ket{J_{RS+EPD}} = \ket{R_{RS+EPD}}, \label{Jvec}
\end{equation}

where $I$ is the identity matrix and

\begin{equation}
\lambda_S^{RS+EPD} = 0+i0, \:\:\:\: \ket{R_{RS+EPD}} =
\begin{pmatrix} \frac{-S_{22}}{S_{21}}\\ 1 \end{pmatrix}. \label{RS_EPD_Evec}
\end{equation}

Solving Equation (\ref{Jvec}) for $\ket{J_{RS+EPD}}$ we find:

\begin{equation}
\ket{J_{RS+EPD}} =
\begin{pmatrix} \frac{1}{S_{21}}\left(1 - bS_{22}\right)\\ b \end{pmatrix} \label{Jordan_vec}
\end{equation}

where $b$ is an arbitrary coefficient. Equation (\ref{Jordan_vec}) and $\ket{R_{EPD}}$, $\ket{R_{RS+EPD}}$ are only correct if $S_{21} \neq 0$. In the limit that $S_{21}$ goes to $0$, the alternate eigenvector formula from supplementary material \ref{sec.CPA_Form} can be used to derive the correct vectors.

\rev{\section{Output Signal Attenuation at Robust Splitting Conditions}\label{sec.RS_Loss}}

\begin{figure*}[htb]
\hspace*{-0.3cm}
\centering
\includegraphics[width=18.2cm]{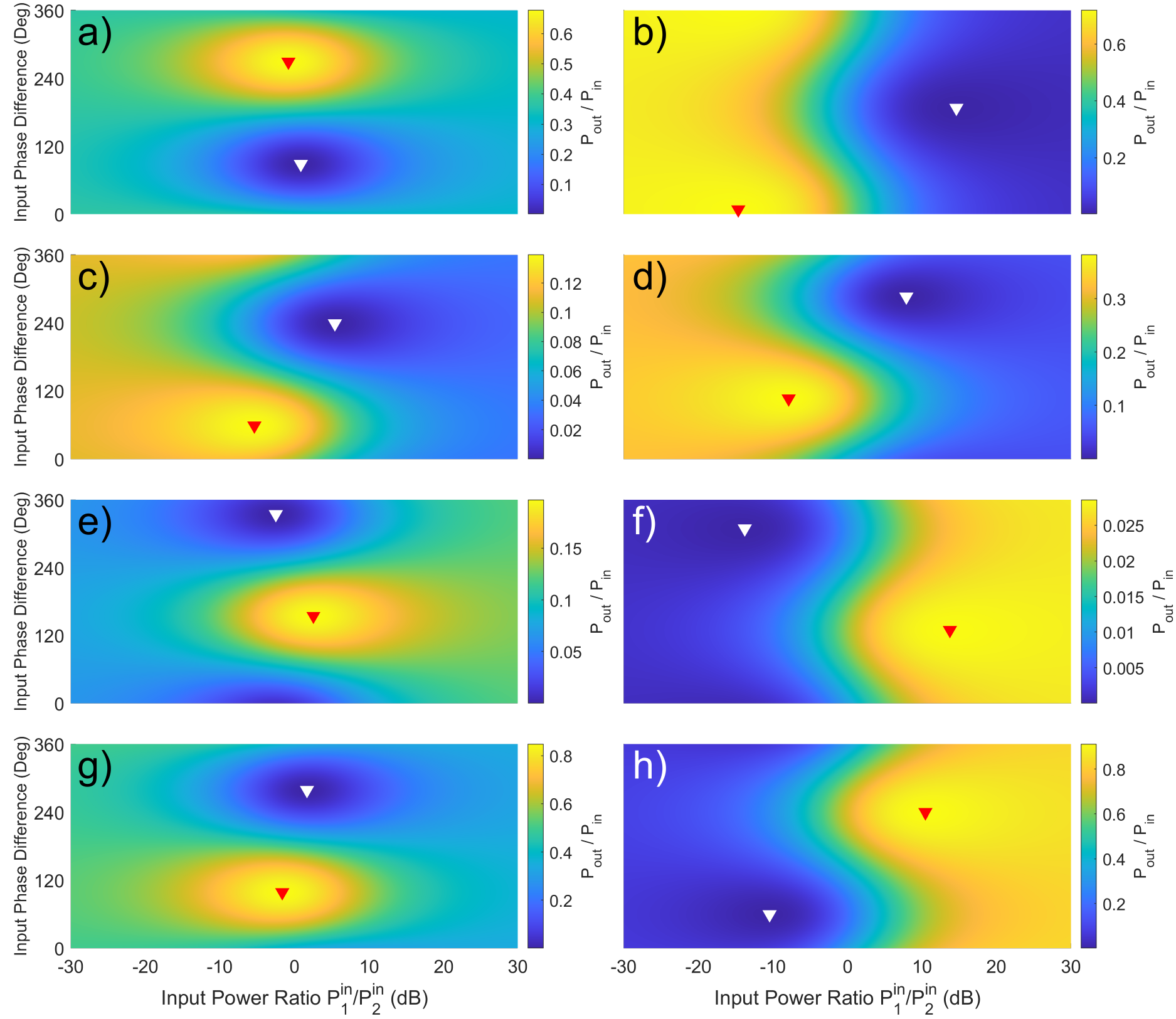}
\caption{\rev{Representative examples of arbitrary signal injection into four different systems set to various RS conditions. In each of the plots, the white triangle indicates the location of the coherent perfect absorption (CPA) eigenvector which corresponds to $\frac{P_{\rm out}}{P_{\rm in}}=0$ and the red triangle indicates the location of the ``Anti-CPA" eigenvector which corresponds to the maximal value of $\frac{P_{\rm out}}{P_{\rm in}}$. Each plot shows $\frac{P_{\rm out}}{P_{\rm in}}$ vs input power ratio and input phase difference for an RS condition measured in a (a-b) reciprocal tetrahedral graph, (c-d) quarter bow-tie two-dimensional microwave billiard, (e-f) three-dimensional microwave cavity, and a (g-h) Random Matrix Theory model. Note that the color scales are different in each panel because each RS condition has a different value of the non-zero eigenvalue.}}
\label{Injection_Loss}
\end{figure*}

\rev{When injecting signals into a generic complex scattering system, the signals will experience some inevitable absorption as they travel through the system. From Equations (\ref{RS_in}) - (\ref{RS_out}), assuming that the first eigenvector corresponds to the zero eigenvalue, we see that the amount of loss that signals injected into systems at RS conditions receive is dependent on $c_1$ and $\lambda_S^{(2)}$. The entire $c_1$ portion of the input signal is completely absorbed since $\ket{R_1}$ corresponds to the coherent perfect absorption (CPA) eigenvector. The amount of loss that $\lambda_S^{(2)}$ imparts on the input signals depends on how far $|\lambda_S^{(2)}|$ is from 1. For unitary scattering systems (systems with no loss/gain), RS conditions aren't possible to achieve and the scattering eigenvalues will live on the unit circle in the complex plane. Adding loss to the system will push the scattering eigenvalues away from the unit circle towards the origin and allow RS conditions to occur. The greater the loss, the more the eigenvalues will tend to move away from the unit circle. Therefore, for lossier systems at RS conditions, the non-zero scattering eigenvalue will be statistically less likely to be near the unit circle, generally imparting more loss on the output signals. 

To visualize the degree of loss, we proceed as follows.  Experimentally, all the information relating the input signals to the output signals is contained in the scattering matrix (Eq. (\ref{S_def})). Therefore, for every RS condition that we find experimentally, with the associated scattering matrix we can construct any input signal and apply the scattering matrix to obtain the output signals. Then we can make a surface plot showing the absorbed power of the output signals compared to the input signals by plotting $\frac{P_{\rm out}}{P_{\rm in}} = \frac{|V_1^{\rm out}|^2+|V_2^{\rm out}|^2}{|V_1^{\rm in}|^2+|V_2^{\rm in}|^2}$ vs input power ratio and input phase difference for various RS conditions. 

For our experimental systems, on average the 1D graph is less lossy than the 2D billiard, and the 2D billiard is less lossy than the 3D cavity. Although for comparable RS conditions within a system, there can be instances where the non-zero eigenvalue is near the unit circle or is near zero. Due to this, to properly determine the behavior of RS conditions on average within a system, a statistical approach is needed and will be presented later in this section. 

In Figure \ref{Injection_Loss} we show surface plots of $\frac{P_{\rm out}}{P_{\rm in}}$ for eight representative examples of RS conditions from three experimental systems, one from each system dimension (1D, 2D, 3D), and the Random Matrix Theory (RMT) model. For the RMT model we set its uniform attenuation rate to be slightly less than our 1D graph. The plots show larger overall attenuation (quantified as $P_{\rm out}/P_{\rm in}$) as blue and smaller attenuation as yellow.  The structure of the $\frac{P_{\rm out}}{P_{\rm in}}$ plots is determined by the relative components of the CPA eigenvector. For RS conditions where the relative amplitudes of the CPA eigenvector components are near one, there are two distinct features corresponding to the locations of the CPA and ``Anti-CPA" eigenvectors (the ``Anti-CPA" eigenvector is not the non-CPA eigenvector $\ket{R_2}$, and will be explained later) similar to what is seen in Figure \ref{Injection_Loss}(a). For RS conditions where the relative amplitudes of the CPA eigenvector components are significantly different than one, the high and low loss regions corresponding to the CPA and ``Anti-CPA" eigenvector locations are stretched out toward the sides of the plot, similar to Figure \ref{Injection_Loss}(b). This occurs  when one component of the CPA eigenvector is significantly different than the other, hence input signal injections near the CPA and ``Anti-CPA" eigenvector locations are therefore mostly single port excitations, so small changes in input power ratio or phase difference don't affect the output signal very much. For input power ratios greater than $|30|$ dB, the output signal is dominated by the larger input vector component and increasing the input power ratio further has a negligible effect on the power of the output signals. 

The location of the point of maximal output power is what we have been calling the ``Anti-CPA" eigenvector location. The ``Anti-CPA" eigenvector can be determined following the method described in Refs. \citenum{Li17,Chen20}. First, calculate the eigenvectors and eigenvalues of the absorption matrix $A \equiv 1_2 - S^\dagger S$, using the scattering matrix at the RS condition. Then, the eigenvector corresponding to the minimum eigenvalue is the ``Anti-CPA" eigenvector, and the CPA eigenvector has an absorption eigenvalue of 1. The ``Anti-CPA" eigenvector or maximal output power state can also be determined using singular value decomposition \cite{Liew2014,Shi2015,Miller2017,Miller2019,Guo2023}. We find that at every RS condition, if the CPA eigenvector has a relative power ratio \textit{P} dB and a phase difference of $\theta$, then the ``Anti-CPA" eigenvector will always have a relative power ratio of -\textit{P} dB and a phase difference of $\theta \pm 180^{\circ}$. This relationship can be seen in all of the $\frac{P_{\rm out}}{P_{\rm in}}$ plots in Figure \ref{Injection_Loss}, as the white triangles (CPA eigenvector locations) and the red triangles (``Anti-CPA" eigenvector locations) are always $180^{\circ}$ apart and have opposite input power ratios.} 

\begin{figure}[htb]
\hspace*{-0.3cm}
\centering
\includegraphics[width=8.7cm]{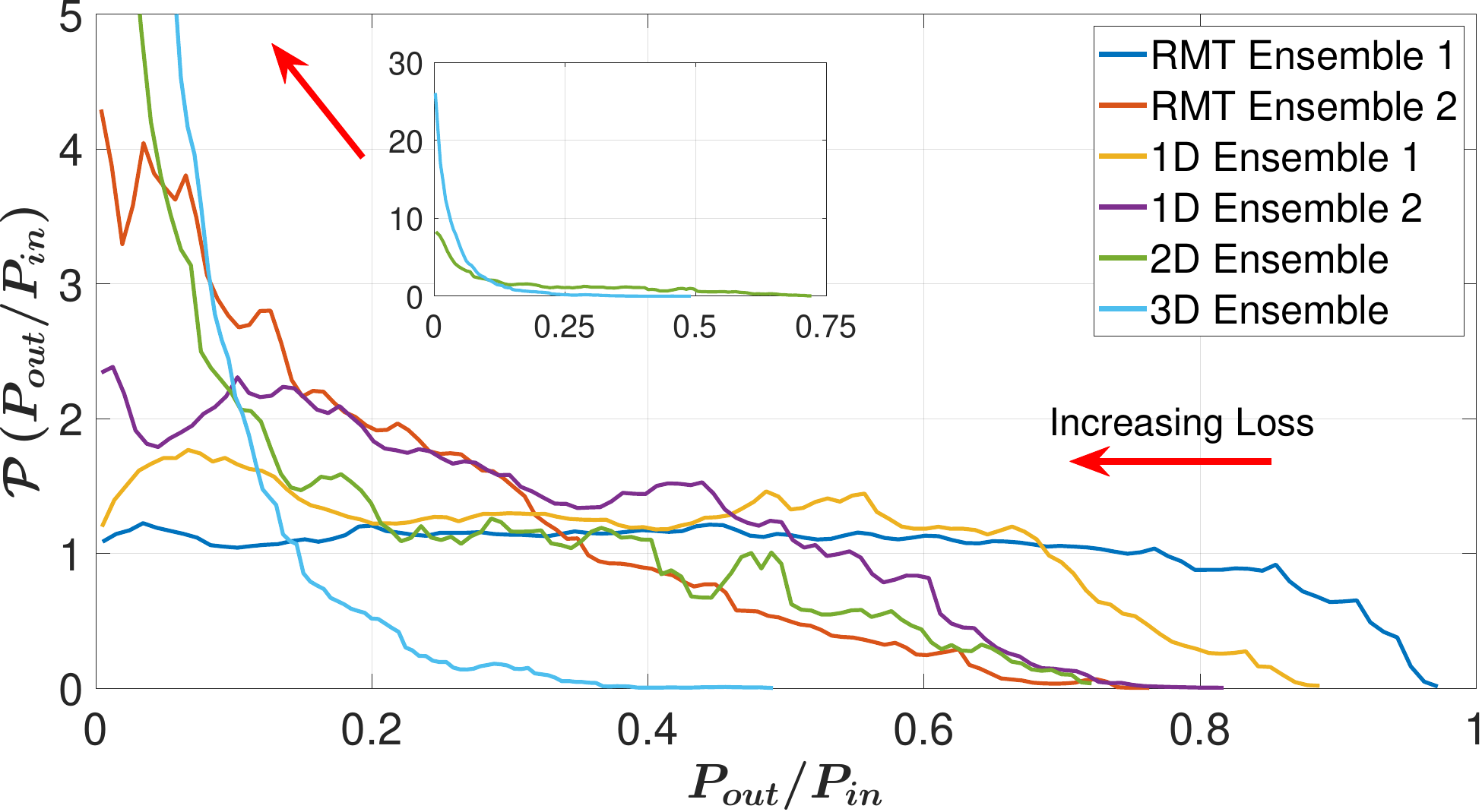}
\caption{\rev{PDFs of $\frac{P_{\rm out}}{P_{\rm in}}$ from six ensembles of reciprocal RS conditions from three experimental systems and the RMT model. The RMT ensembles were created following the details of supplementary material \ref{sec.RMT}, the experimental 1D ensembles were from reciprocal tetrahedral graphs, the experimental 2D ensemble was from the quarter bow-tie two-dimensional microwave billiard, and the experimental 3D ensemble was from the three-dimensional microwave cavity. The inset shows the full PDF of the 2D and 3D ensembles.}}
\label{Loss_PDF}
\end{figure}

\rev{To determine the probability distribution of absorbed power that the output signals have for various RS conditions in our systems, we need to first collect a significant number of comparable RS conditions for each system. To do this, instead of parametrically measuring the scattering matrix vs two tunable parameters as we have discussed previously, we measure the scattering matrix vs frequency (in a reasonably small frequency range) for many different settings of the tunable devices within a system. Each realization of $S$ vs frequency should be statistically independent of each other. Then, we analyze each realization and look for all scattering matrices where $|det(S)|$ is below some threshold (usually 0.004), corresponding to an RS condition. If there are multiple points in succession where $|det(S)|$ is below the threshold, we only include the scattering matrix with the smallest value of $|det(S)|$. Next, for all of the scattering matrices (or RS conditions) we have collected over the ensemble, we apply the scattering matrix of each RS condition to input signals with 1001 power ratios between -30 and +30 dB and 1001 phase differences between $0^{\circ}$ and $360^{\circ}$ to obtain the output signals. We don't go beyond input power ratios of $|30|$ dB as the input signals beyond this range are strongly dominated by a single component (port), causing minimal change in the output power. Then we calculate $\frac{P_{\rm out}}{P_{\rm in}}$ for all 1001x1001 input signals, similar to the surface plots shown in Figure \ref{Injection_Loss}. Lastly, we combine all of the $\frac{P_{\rm out}}{P_{\rm in}}$ data for every RS condition in the ensemble and calculate the probability density function (PDF) of $\frac{P_{\rm out}}{P_{\rm in}}$. 

In Figure \ref{Loss_PDF}, we show the PDFs of $\frac{P_{\rm out}}{P_{\rm in}}$ from six ensembles of reciprocal RS conditions from three experimental systems, as well as the RMT model. In RMT Ensemble 1 there were 124 RS conditions, the 1D Ensemble 1 had 100 RS conditions, the 1D Ensemble 2 had 194 RS conditions, the RMT Ensemble 2 had 213 RS conditions, the 2D Ensemble had 267 RS conditions, and the 3D Ensemble had 1587 RS conditions. The ensembles ordered from least lossy to most lossy on average are: RMT Ensemble 1, 1D Ensemble 1, 1D Ensemble 2, RMT Ensemble 2, 2D Ensemble, and 3D Ensemble. In Figure \ref{Loss_PDF}, we see that the lower loss ensembles follow a more uniform shape until falling off at high values of $\frac{P_{\rm out}}{P_{\rm in}}$, while the higher loss ensembles have a strong decay away from $\frac{P_{\rm out}}{P_{\rm in}}=0$ and are significantly less likely to reach high values of $\frac{P_{\rm out}}{P_{\rm in}}$.}

\rev{\section{Form of Scattering Eigenvectors at Robust Splitting Conditions} \label{sec.CPA_Form}}
For a $2\times 2$ scattering matrix, the most general way to write the eigenvalues and eigenvectors is:

\begin{equation}
\lambda_S^{(1,2)} = \frac{S_{11}+S_{22}}{2} \pm \frac{1}{2} \sqrt{(S_{11} - S_{22})^2 + 4 S_{12} S_{21}}
\end{equation}

\begin{equation}
 \ket{R_{1,2}} =
\begin{pmatrix}
\frac{1}{S_{21}}\left(\lambda_S^{(1,2)} - S_{22}\right)\\
1
\end{pmatrix}.
\end{equation}

Note that in the limit that $S_{21}$ goes to $0$, the eigenvectors should be of the form $\ket{R_{1,2}} = \begin{pmatrix} 1 \\ \frac{1}{S_{12}}\left(\lambda_S^{(1,2)} - S_{11}\right) \end{pmatrix}$. In the limit that $S_{21}$ and $S_{12}$ go to $0$, the system is trivially diagonalized, has an orthonormal basis, and EPD’s are not possible.

If a system is set to \rev{RS} conditions, there are generally three possibilities that the output signals can take. For all cases below, in the limit that $S_{21}$ goes to $0$ the equations are incorrect, but the correct results can be derived using the alternate eigenvector formula specified above.

\subsection{$\lambda_S^{(1)}=0+i0$}

\begin{equation}
 \ket{R_{1}} = 
\begin{pmatrix}
\frac{-S_{22}}{S_{21}}\\
1
\end{pmatrix}, \:\:\:\:\:
\ket{R_{2}} = 
\begin{pmatrix}
\frac{1}{S_{21}}\left(\lambda_S^{(2)} - S_{22}\right)\\
1
\end{pmatrix}
\end{equation}

\begin{equation}
\genfrac{(}{)}{0 pt}{}{V_1^{\rm in}}{V_2^{\rm in}} = c_1 \ket{R_{1}} + c_2 \ket{R_{2}}
\end{equation}

\begin{equation}
\genfrac{(}{)}{0 pt}{}{V_1^{\rm out}}{V_2^{\rm out}} = S \genfrac{(}{)}{0 pt}{}{V_1^{\rm in}}{V_2^{\rm in}} = c_2 \lambda_S^{(2)} \ket{R_{2}}
\end{equation}

\subsection{$\lambda_S^{(2)}=0+i0$}

\begin{equation}
 \ket{R_{2}} = 
\begin{pmatrix}
\frac{-S_{22}}{S_{21}}\\
1
\end{pmatrix},\:\:\:\:\:
 \ket{R_{1}} = 
\begin{pmatrix}
\frac{1}{S_{21}}\left(\lambda_S^{(1)} - S_{22}\right)\\
1
\end{pmatrix}
\end{equation}

\begin{equation}
\genfrac{(}{)}{0 pt}{}{V_1^{\rm in}}{V_2^{\rm in}} = c_1 \ket{R_{1}} + c_2 \ket{R_{2}}
\end{equation}

\begin{equation}
\genfrac{(}{)}{0 pt}{}{V_1^{\rm out}}{V_2^{\rm out}} = S \genfrac{(}{)}{0 pt}{}{V_1^{\rm in}}{V_2^{\rm in}} = c_1 \lambda_S^{(1)} \ket{R_{1}}
\end{equation}

\subsection{$\lambda_S^{(1)}=\lambda_S^{(2)}=\lambda_S^{RS+EPD}=0+i0$}
For a detailed explanation of the input and output signal derivation, $\ket{J_{RS+EPD}}$, and why the output signal isn't $0$ even though both eigenvalues are $0+i0$, see supplementary material \ref{sec.Jordan}.

\begin{equation}
 \ket{R_{1}} = \ket{R_{2}} = \ket{R_{RS+EPD}} = 
\begin{pmatrix} \frac{-S_{22}}{S_{21}}\\ 1 \end{pmatrix}
\end{equation}

\begin{equation}
\ket{J_{RS+EPD}} =
\begin{pmatrix} \frac{1}{S_{21}}\left(1 - bS_{22}\right)\\ b \end{pmatrix},
\end{equation}

where $b$ is an arbitrary coefficient.

\begin{equation}
\genfrac{(}{)}{0 pt}{}{V_1^{\rm in}}{V_2^{\rm in}} = c_1 \ket{R_{RS+EPD}} + c_2 \ket{J_{RS+EPD}}
\end{equation}

\begin{equation}
\genfrac{(}{)}{0 pt}{}{V_1^{\rm out}}{V_2^{\rm out}} = S \genfrac{(}{)}{0 pt}{}{V_1^{\rm in}}{V_2^{\rm in}} = c_2 \ket{R_{RS+EPD}}
\end{equation}

\section{Signal Injection at Arbitrary Conditions} \label{sec.Arbitrary_Inj}

\begin{figure}[htb]
\centering
\includegraphics[width=8.7cm]{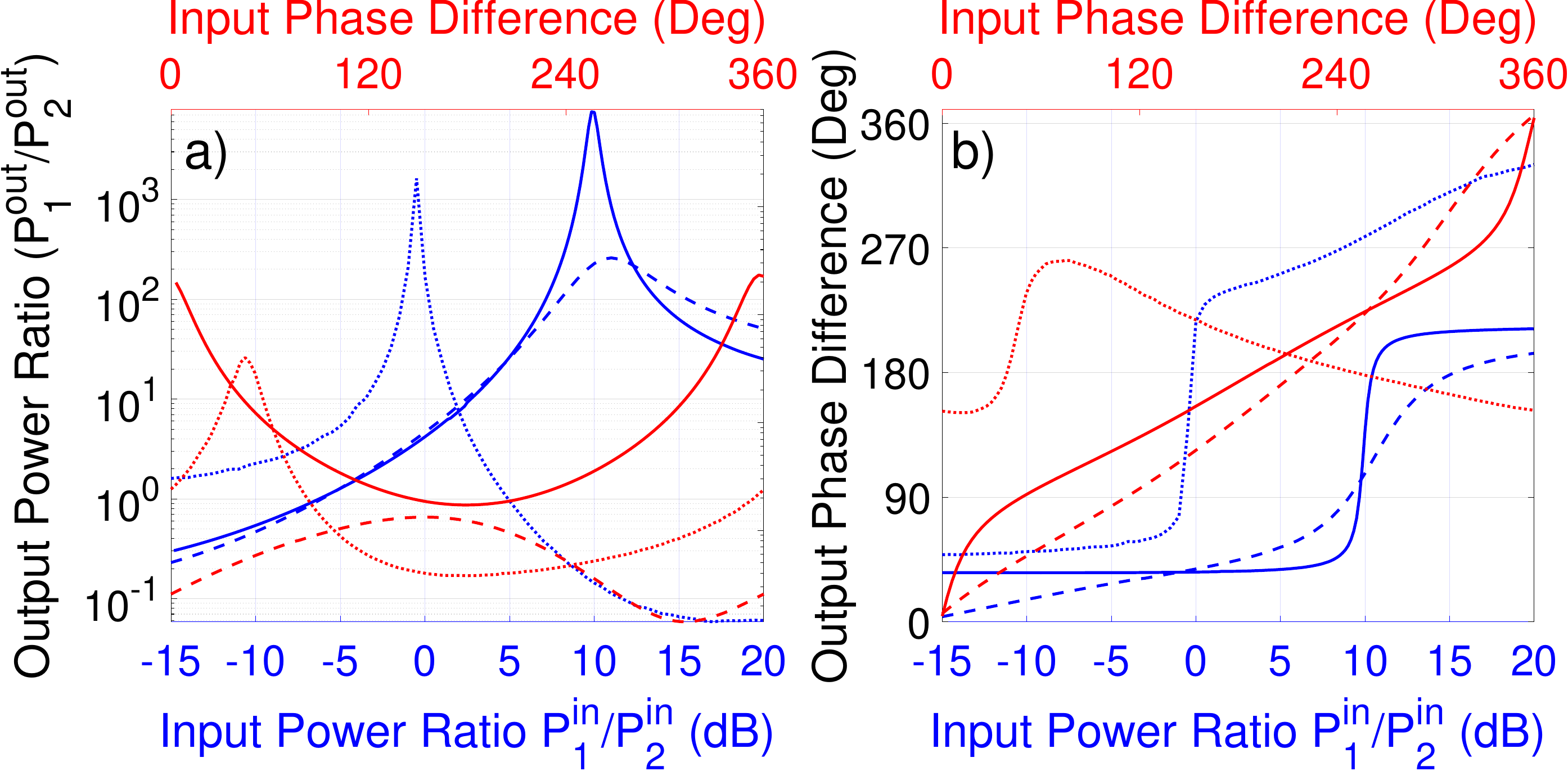}
\caption{Experimental demonstration of arbitrary signals injected into three different experimental systems away from \rev{RS} conditions. a) Output power ratio vs input power ratio (lower axis, blue) and input phase difference (upper axis, red). The blue curves correspond to the input signals power ratio being swept, and the red curves correspond to the input signals relative phase being swept. The solid curves for each color denote measurements in a one-dimensional graph, the long dashed curves denote measurements in a two-dimensional billiard, and the short dashed curved denote measurements in a three-dimensional cavity. b) Output phase difference vs input power ratio (lower axis, blue) and input phase difference (upper axis, red). The blue and red curves (solid and dashed) have the same interpretation as in (a). These results contrast sharply with those shown in Fig. \ref{Multiple_CPA}.}
\label{Arbitrary_Injection}
\end{figure}

In this work, we have focused on signal injection into systems set to \rev{RS} conditions. Here, in Figure \ref{Arbitrary_Injection} we show representative examples of arbitrary signals injected into three different systems not at an \rev{RS} condition to contrast with the robust splitting phenomenon at systems set to \rev{RS} conditions. For these arbitrary signals, we chose the settings of the phase shifters/metasurfaces and frequency to be a random value before signal injection. Since the systems are not set to any unique conditions, the output power ratio and phase difference between the two ports varies dramatically as the input signals change. For clarity, we have adjusted some of the output phase differences by an overall constant.

\section{Dynamics of Robust Splitting Conditions} \label{sec.Movie}
\begin{figure}[htb]
\hspace*{-0.28cm}
\centering
\includegraphics[width=9.1cm]{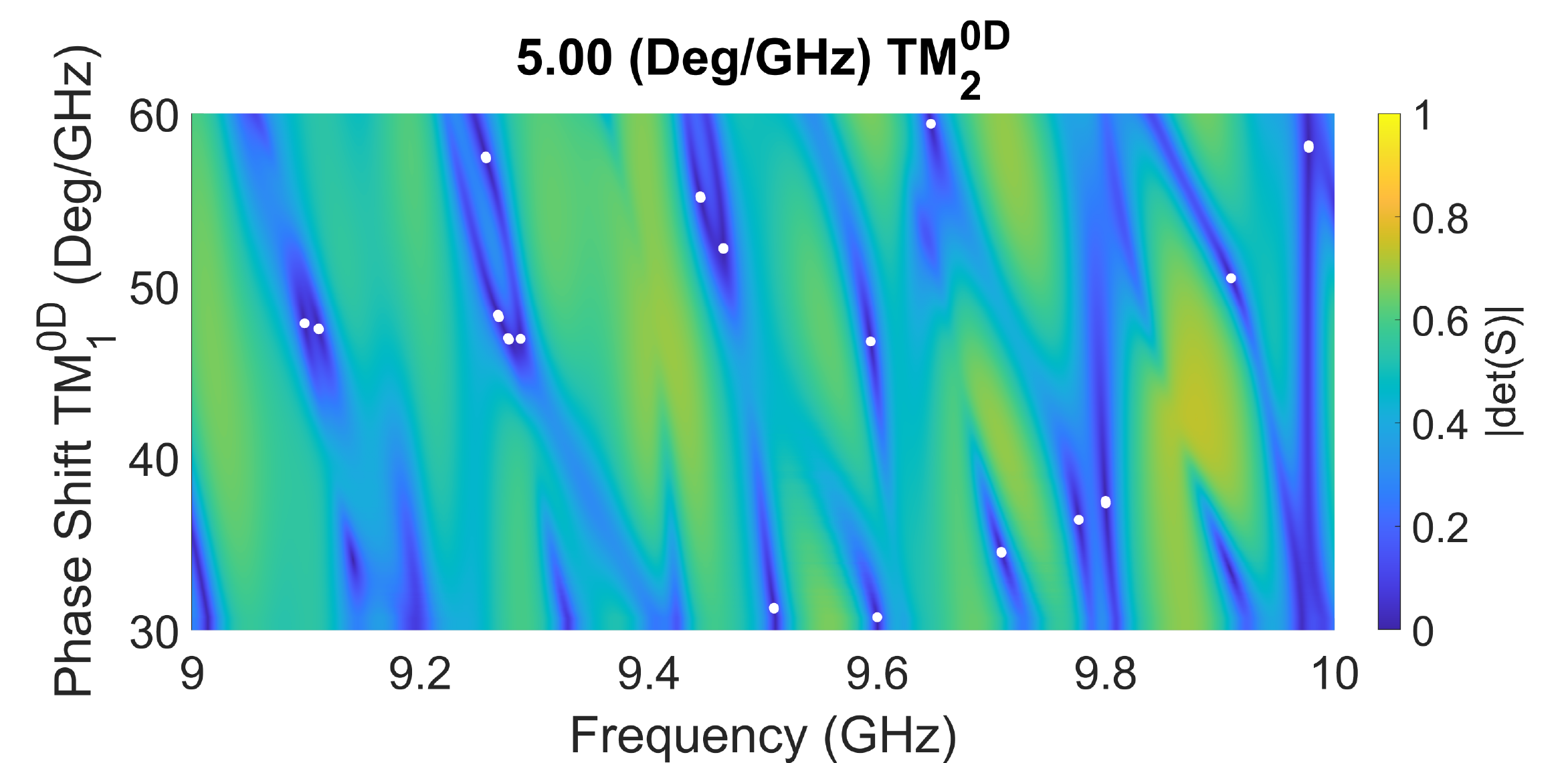}
\caption{Dynamics and interactions of \rev{robust splitting} conditions as a third parameter of the system is varied. The image above shows the first frame of a video which explores the dynamics of \rev{RS} conditions in the reciprocal tetrahedral microwave graph, including the \rev{RS} condition shown in Figure \ref{Tunable_CPA}a. Plotted is the $|det(S)|$ vs frequency and phase shift of $TM_1^{0D}$ and each frame of the video is at a different fixed phase shift of $TM_2^{0D}$. In the video the \rev{RS} conditions are highlighted by the white points where $|det(S)| \le 0.002$, and we see numerous creation and annihilation events of \rev{RS} conditions as the phase shift of $TM_2^{0D}$ varies. (Multimedia available online)}
\label{Video_1}
\end{figure}

In Figure \ref{Tunable_CPA}a, the motion of a single \rev{RS} condition as the third parameter of the system varied was shown. In Figure \ref{Video_1}, we show a much broader view of $|det(S)|$ vs frequency and phase shift of $TM_1^{0D}$, showing the dynamics and evolution of the \rev{RS} conditions over a large sweep of the third parameter ($TM_2^{0D}$), including many instances of pair creation and annihilation events (Multimedia available online). 

\rev{\section{Stability of Robust Splitting Conditions Over Time} \label{sec.Stability}}
Any parameter that can change the boundary conditions or internal geometry of a system can have a significant effect on the scattering properties of the system. We purposefully use phase shifters and metasurfaces as tunable parameters to manipulate the scattering singularities, but once a system is set at conditions of a scattering singularity, such as \rev{RS} conditions, how stable are these singularities? Once the controllable parameters of the system are set (such as frequency, metasurface applied bias voltage, etc.), as time progresses the environment in which the physical system is in can slowly change. 

\rev{In general, it takes time to parametrically measure the scattering matrix and identify a particular RS condition. It takes additional time to convert the VNA into the 2-port dual-source mode and inject various signals into the system. During the time after the RS condition has been identified, it can drift away from its original location. When this occurs, one of the eigenvectors no longer has a zero eigenvalue, so both eigenvectors contribute to the output signal, causing the output power ratio and phase difference to vary for different input signals. Our 3D system is by far the most sensitive to changes in the environment, making it the most likely to have deviations from the ideal flat response, as seen by the green curves in Fig. \ref{Multiple_CPA}. Moving faster from identifying the RS conditions to injecting various signals would reduce these deviations, for short term measurements.

There are other methods that could be implemented to reduce the effects of longer-term environmental change on experimental systems. The simplest method would be to have the experimental system in a temperature-controlled environment, although this is not always feasible in real-world environments. A more sophisticated method would be to implement a feedback loop to automatically adjust the values of the tunable parameters of a system to keep an RS condition steady over time at a fixed frequency. Experimental systems can also be carefully designed to be insensitive to environmental fluctuations. In particular, small systems (such as graphs with short overall lengths) are significantly less affected by environmental changes. Additionally, with a specifically engineered system, it's likely possible to know the parameter values of the tunable devices required to produce the RS conditions. This is advantageous because due to the complexity (and environmental time dependence) of generic experimental systems, the required system parameters for RS conditions can't be precomputed.}

\begin{figure}[htb]
\centering
\includegraphics[width=8.7cm]{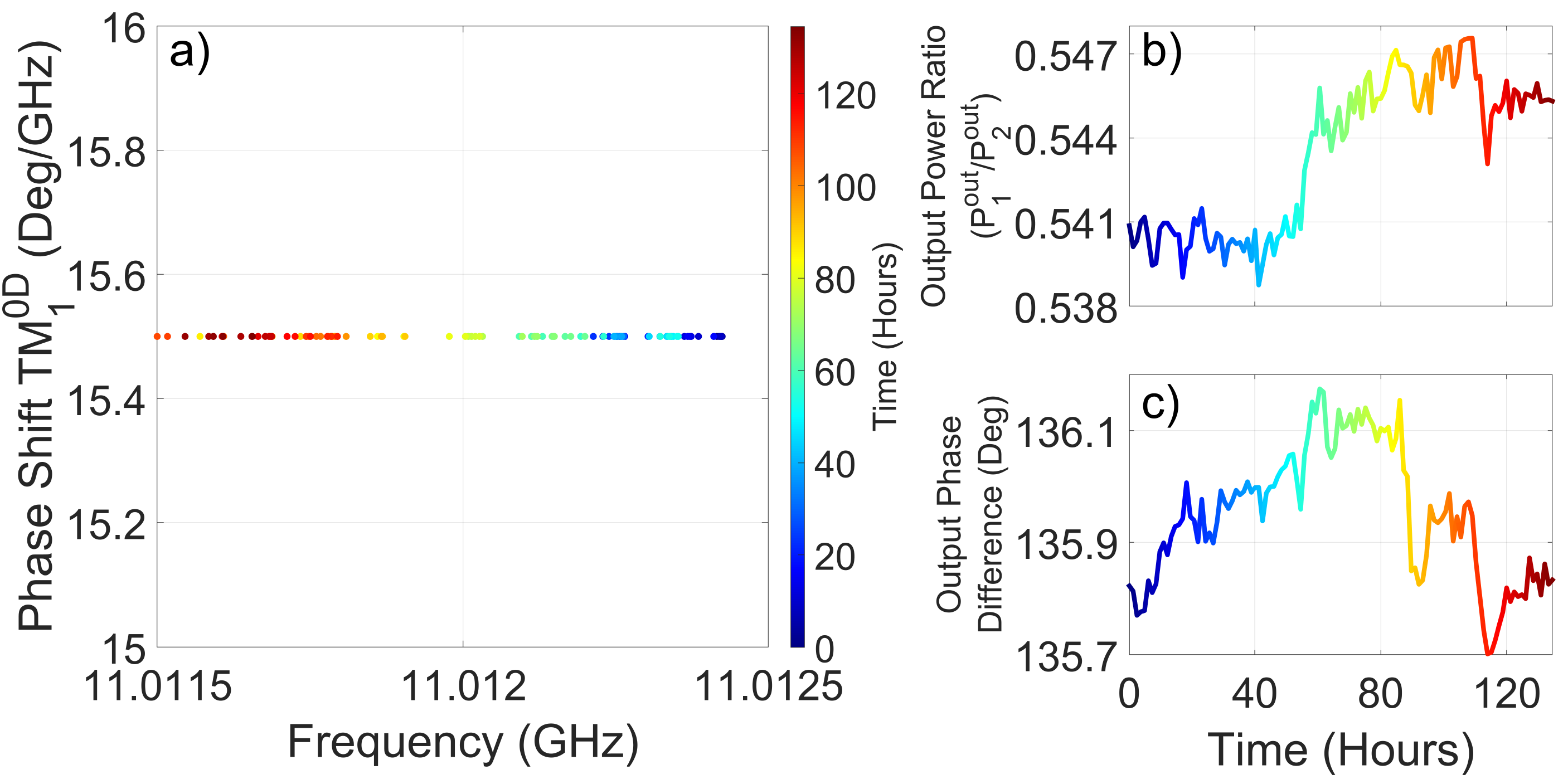}
\caption{Experimental demonstration of the stability of \rev{robust splitting} conditions in an approximately 8.2m long 6-node bipartite microwave graph. The color of each point corresponds to the time the \rev{RS} condition was measured at. (a) Location of an \rev{RS} condition in the two-dimensional parameter space of frequency and phase shift of $TM_1^{0D}$ over time. (b) Output power ratio vs time. (c) Output phase difference vs time, over 130 hours.}
\label{Fixed_CPA_Graph}
\end{figure}

\begin{figure}[htb]
\centering
\includegraphics[width=8.7cm]{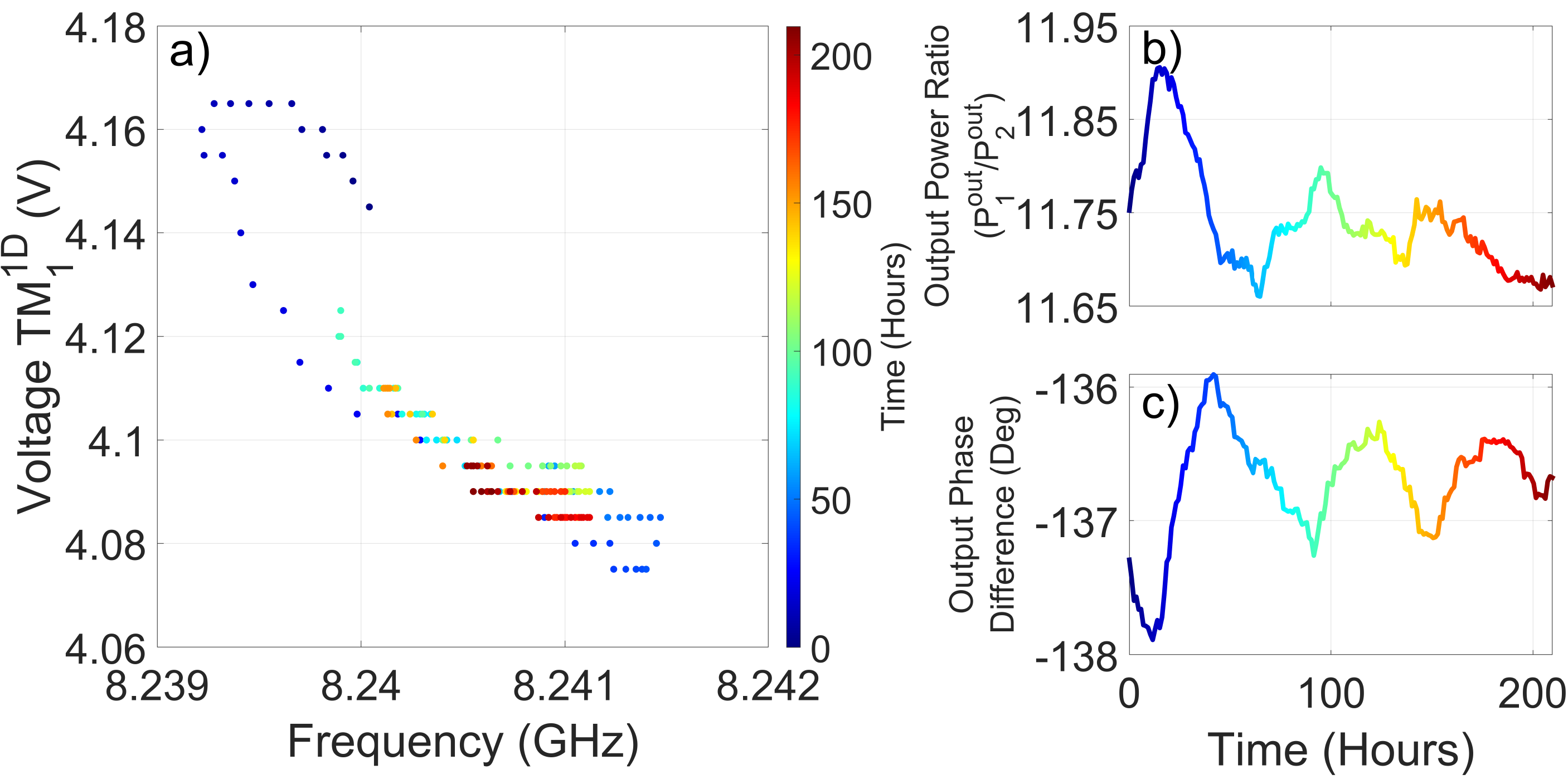}
\caption{Experimental demonstration of the stability of \rev{robust splitting} conditions in the quarter bow-tie two-dimensional billiard. The color of each point corresponds to the time the \rev{RS} condition was measured at. (a) Location of an \rev{RS} condition in the two-dimensional parameter space of frequency and $TM_1^{1D}$ applied bias voltage over time. (b) Output power ratio vs time. (c) Output phase difference vs time, over 200 hours.}
\label{Fixed_CPA_Billiard}
\end{figure}

In this work, the largest change in the environment is the temperature fluctuation of the laboratory. In general, the temperature of our environment can fluctuate by 6 degrees Fahrenheit per day due to \rev{external weather fluctuations as well as} the heating and/or cooling schedule. \rev{Additionally, the heating and/or cooling schedule is not consistent throughout each week.} This has a perturbative effect on the size and boundary conditions of a system, as the change in temperature will expand or contract the materials and/or air volume comprising the structure of the system. The magnitude of the perturbation scales with the size of the system. To quantify this change over time, we measured the same two-dimensional parameter space which had a single \rev{RS} condition over the course of multiple days in both a microwave graph and a billiard (Figs. \ref{Fixed_CPA_Graph}, \ref{Fixed_CPA_Billiard}). 

In Figure \ref{Fixed_CPA_Graph}, measured in an 8.2m long microwave graph, the left plot shows the location of an \rev{RS} condition varying in the frequency and phase shift of $TM_1^{0D}$ parameter space over time. The resolution of $TM_1^{0D}$ in this plot was 0.02 Deg/GHz, but the \rev{RS} condition moved minimally in this dimension, so all points have the same y-value of 15.5 Deg/GHz. The right plots show how the output power ratio and phase difference vary along the drifting trajectory of the \rev{RS} condition. The colors show the evolution over time. There is very little change in the location of the \rev{RS} condition and in the output power ratio and phase difference over time, showing the strong stability of graphs. 

In Figure \ref{Fixed_CPA_Billiard}, measured in the quarter bow-tie two-dimensional billiard, we show similar results over time as in Figure \ref{Fixed_CPA_Graph}. But in the two-dimensional billiard, the location of the \rev{RS} condition and the output power ratio and phase difference are perturbed much more significantly than in the graph. Although, overall the location and output values are fairly stable over the almost 9 days of measurements. For the three-dimensional microwave cavity, the locations of \rev{RS} conditions and their output signals were extremely sensitive to environmental changes over time. Even over a relatively short period of time, the system drifts away from \rev{RS} conditions, as seen by the fluctuating green curves in Figure \ref{Multiple_CPA}.

\section{Random Matrix Numerics Modeling}\label{sec.RMT}
The theoretical modeling of wave scattering in this section follows the details described in Ref. \citenum{Erb25}. The scattering matrix $S$ that connects incoming with outgoing waves $|\alpha_{out}\rangle=S|\alpha_{in}\rangle$, takes the form:

\begin{equation}
\begin{aligned}
S(\omega) &= -1_2 + i W G(\omega)W^T,\\ 
G(\omega) &= \frac{1}{\omega - H_{eff}}.
\end{aligned}
\end{equation}

where $\omega$ is frequency, $W$ is the $2\times \mathcal{N}$ coupling matrix, $H_{eff} = H_0 - \frac{i}{2} W^T W$ is the effective Hamiltonian, and $H_0$ is the Hamiltonian that describes the isolated system \cite{Erb25}. 

The Hamiltonian $H_0$ is modeled by a random matrix ensemble when the scattering system is a cavity with underlying chaotic ray dynamics. This ensemble is the Gaussian Orthogonal Ensemble (GOE) for time-reversal symmetric (TRS) systems, while for systems where TRS is violated the appropriate ensemble is the Gaussian Unitary Ensemble \cite{Stock99,Haake10}. To model systems with additional parametric degrees of freedom ($x,y$), $H_0$ takes the form:

\begin{equation}
H_0(x,y) = H_1 + \lvert cos(x) \rvert \lvert cos(y) \rvert H_2 + \lvert sin(x) \rvert \lvert sin(y) \rvert H_3.
\label{eqn:H0Def2}
\end{equation}

The elements of $H_1$ are taken from a GOE of mean zero and standard deviation $\sqrt{\frac{\mathcal{N}}{\pi}}$. The diagonal elements of $H_2$ and $H_3$ are taken from the uniform distribution $[0,-i]$, while the off-diagonal elements are taken from a GOE of the same family as $H_1$. The Hamiltonian $H_1$ models unperturbed chaotic cavities, while the non-Hermitian Hamiltonians $H_{2,3}$ model the additional degrees of freedom, such as metasurfaces, which perturb the cavities.

To model a system with non-reciprocity, we introduce a magnetic field:
\begin{equation}
H_{mag} = H_0 + i \alpha B,
\end{equation}
where $B$ is an anti-symmetric GOE matrix and $\alpha$ controls the relative strength of the magnetic field.

\section{Random Matrix Numerics Results} \label{sec.RMT_Results}
\begin{figure}[htb]
\hspace*{-0.47cm}
\centering
\includegraphics[width=9.3cm]{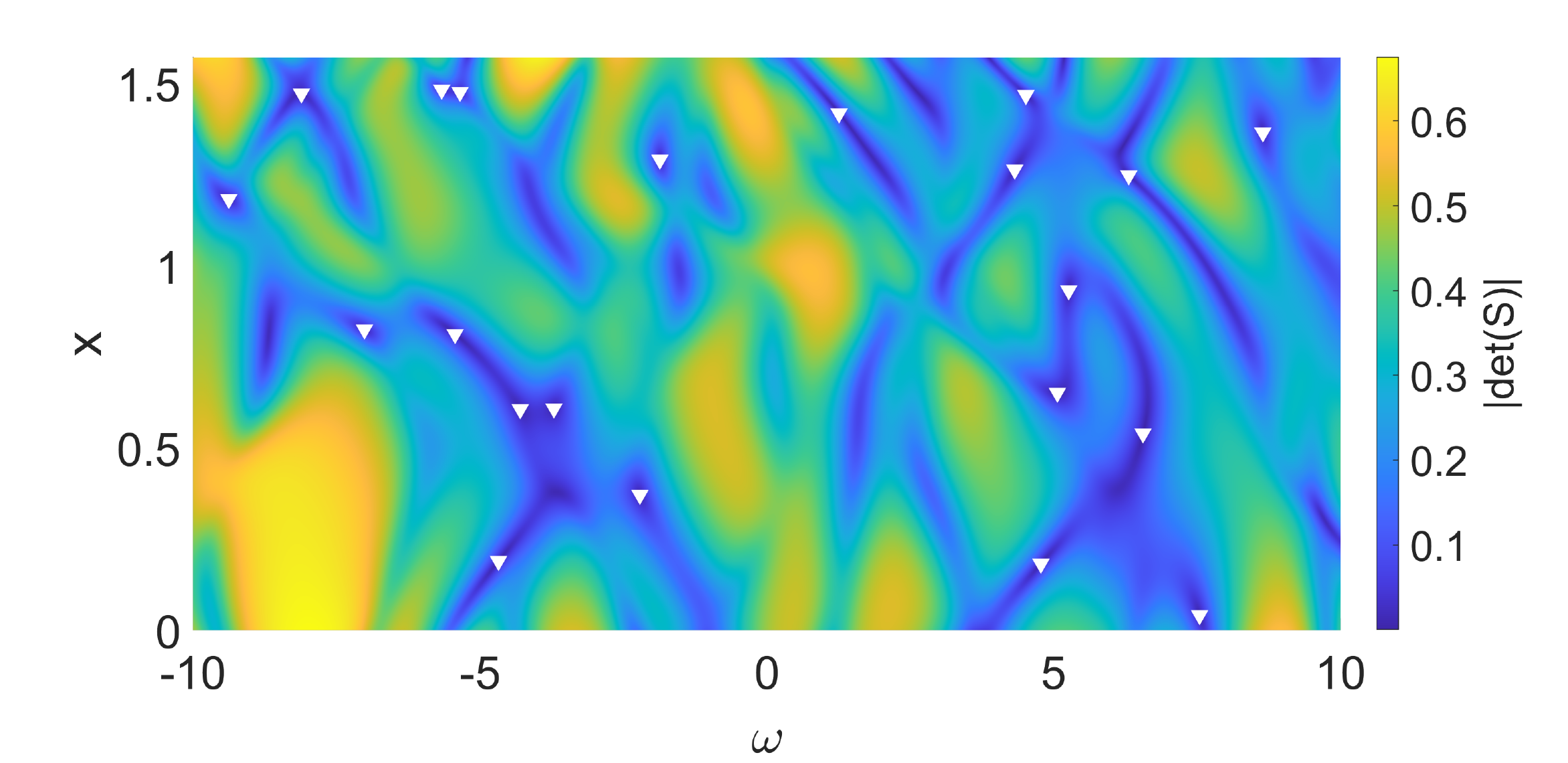}
\caption{Plot of $|det(S)|$ vs $\omega$ and $x$ from the RMT cavity model. The white triangles correspond to points where $det(S)=0+i0$, which enable \rev{robust splitting}.}
\label{Det_S_RMT}
\end{figure}

\begin{figure}[htb]
\centering
\includegraphics[width=8.7cm]{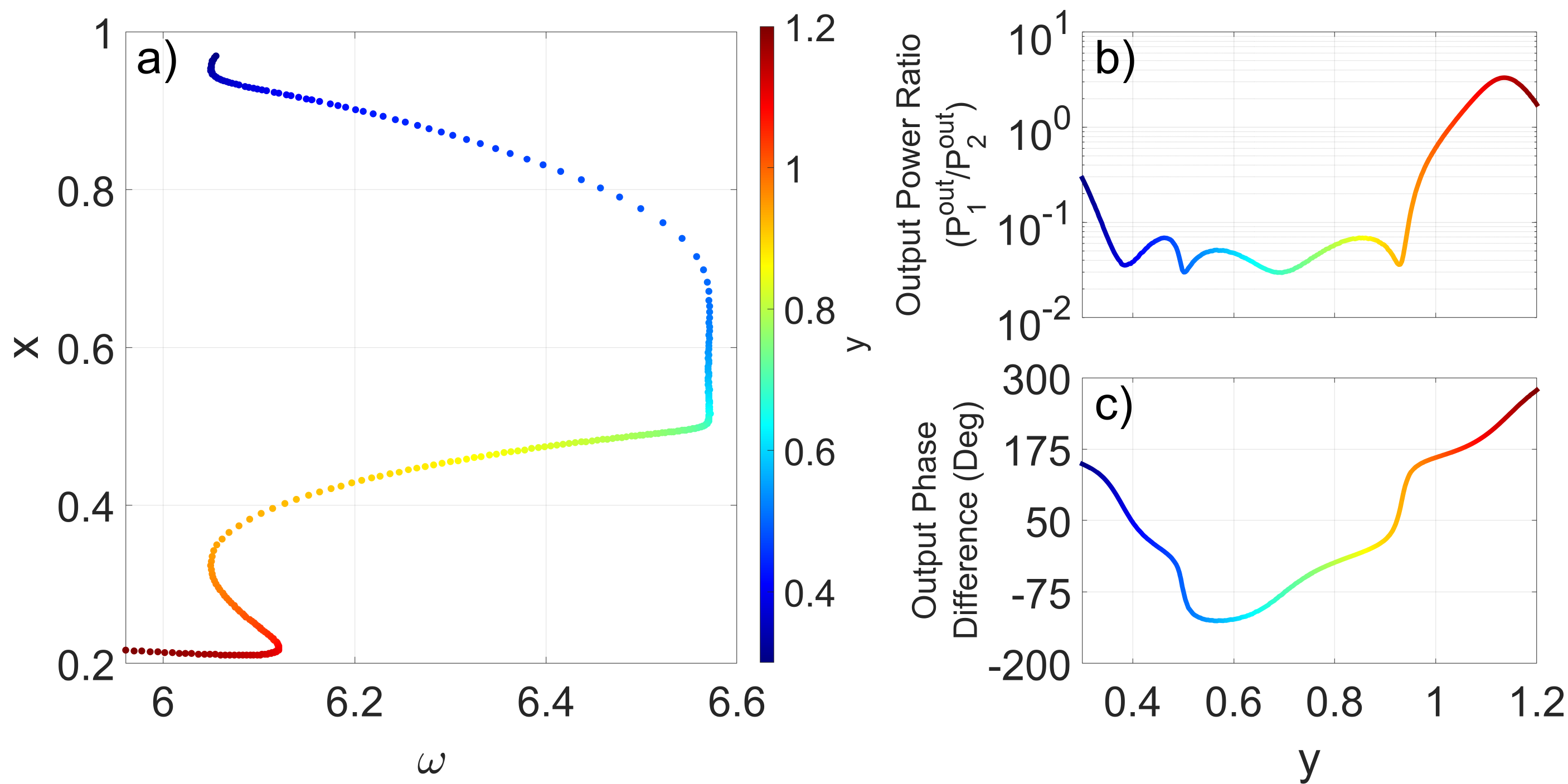}
\caption{Theoretical demonstration of the tunability of \rev{robust splitting} conditions from the RMT cavity model. The color of each point corresponds to the value of the varying third parameter of the model ($y$) the \rev{RS} condition was simulated at. (a) Location of an \rev{RS} condition in the two-dimensional parameter space of $\omega$ and $x$ as $y$ varies. (b) Output power ratio vs $y$. (c) Output phase difference vs $y$.}
\label{Tunable_CPA_RMT}
\end{figure}

To verify the generality of the results measured in physical systems to all wave-scattering systems, we generated data using the RMT model described in supplementary material \ref{sec.RMT} to plot similar results as shown in Figures \ref{Det_S}, \ref{Tunable_CPA} of the main text. Figures \ref{Det_S_RMT}, \ref{Tunable_CPA_RMT} validate that there is an abundance of \rev{RS} conditions that can be manipulated and used as tunable robust splitters in generic wave-scattering systems.

\rev{Note that RMT models include just the bare minimum information to describe a complex scattering system, leaving out details such as geometrical structure or symmetries.  We use it only to test whether certain features in our data are reproduced by this minimal generic model.  It does not make sense to use an RMT model to reproduce the exact detailed properties of any physical system.}

\rev{\section{Signal Injection in Systems with Many Ports}\label{sec.Jordan_Higher}}
In this section, we will describe signal injection and input/output signals in higher dimensional scattering systems, mainly focusing on the three-port case. In a two-port system, there is only one way to achieve an EPD, which is having both eigenvalues and eigenvectors be degenerate. But in systems with more ports, all eigenvalues and eigenvectors, or a subset of them, can be degenerate, and these different cases affect the input and output signals. For an EPD of an $M\ge2$ port system, in the case that $2 \le L \le M$ eigenvalues and eigenvectors are degenerate, we define that EPD to be of order $L$ (EPD-$L$).

Except in the cases of exceptional point degeneracies, which are described later, any signal sent to, and returned from, an arbitrary three-port scattering system can be written as a linear combination of the three $S$-matrix eigenvectors:

\begin{equation}
\begin{pmatrix}V_1^{\rm in}\\V_2^{\rm in}\\V_3^{\rm in} \end{pmatrix} = c_1 \ket{R_{1}} + c_2 \ket{R_{2}} + c_3 \ket{R_{3}}
\end{equation}

\begin{equation}
\begin{pmatrix}V_1^{\rm out}\\V_2^{\rm out}\\V_3^{\rm out} \end{pmatrix} = S \begin{pmatrix}V_1^{\rm in}\\V_2^{\rm in}\\V_3^{\rm in} \end{pmatrix} = c_1 \lambda_S^{(1)} \ket{R_{1}} + c_2 \lambda_S^{(2)} \ket{R_{2}} + c_3 \lambda_S^{(3)} \ket{R_{3}},
\end{equation}

where $\begin{pmatrix}V_1^{\rm in}\\V_2^{\rm in}\\V_3^{\rm in} \end{pmatrix}$, $\begin{pmatrix}V_1^{\rm out}\\V_2^{\rm out}\\V_3^{\rm out} \end{pmatrix}$ are the input and output signals respectively, $\ket{R_{1-3}}$ are the right eigenvectors of the scattering matrix $S$, $c_{1-3}$ are arbitrary coefficients, and $\lambda_S^{(1-3)}$ are the corresponding eigenvalues. In the case where the system is set to \rev{$det(S)=0+i0$} conditions (assume $\lambda_S^{(1)}=0+i0$),

\begin{equation}
\begin{pmatrix}V_1^{\rm out}\\V_2^{\rm out}\\V_3^{\rm out} \end{pmatrix} = c_2 \lambda_S^{(2)} \ket{R_{2}} + c_3 \lambda_S^{(3)} \ket{R_{3}}.
\end{equation}

If the system is set to EPD-2 conditions (assume $\lambda_S^{(1)}=\lambda_S^{(2)}$), we need to complete the eigenbasis using a Jordan vector, following the method described in supplementary material \ref{sec.Jordan} but using a three-port scattering system. Injecting arbitrary signals into a three-port system at EPD-2 conditions, the input and output signals become: 

\begin{equation}
\begin{pmatrix}V_1^{\rm in}\\V_2^{\rm in}\\V_3^{\rm in} \end{pmatrix} = c_1 \ket{R_{EPD-2}} + c_2 \ket{J_{EPD-2}} + c_3 \ket{R_{3}}
\end{equation}

\begin{equation}
\begin{split}
\begin{pmatrix}V_1^{\rm out}\\V_2^{\rm out}\\V_3^{\rm out} \end{pmatrix} = \: &\lambda_S^{EPD-2} \left[c_1 \ket{R_{EPD-2}} + c_2 \ket{J_{EPD-2}}\right] + c_2 \ket{R_{EPD-2}} \\  & + c_3 \lambda_S^{(3)} \ket{R_{3}},
\end{split}
\end{equation}

where $\lambda_S^{EPD-2}$, $\ket{R_{EPD-2}}$ are the degenerate eigenvalues and eigenvectors respectively, and $\ket{J_{EPD-2}}$ is the Jordan vector. If the system is at both \rev{$det(S)=0+i0$} and EPD-2 conditions ($\lambda_S^{EPD-2}=0+i0$), the output signal becomes:

\begin{equation}
\begin{pmatrix}V_1^{\rm out}\\V_2^{\rm out}\\V_3^{\rm out} \end{pmatrix} = c_2 \ket{R_{EPD-2}}  + c_3 \lambda_S^{(3)} \ket{R_{3}}.
\end{equation}

Therefore in a three-port system, a robust output signal doesn't occur in general, even if the system is set to \rev{$det(S)=0+i0$} conditions or EPD-2 and \rev{$det(S)=0+i0$} conditions, as the output signals always depend on at least two vectors.

To describe what happens when a system is set to EPD-3 conditions, the results in supplementary material \ref{sec.Jordan} need to be generalized to higher dimensions of the scattering matrix. When multiple eigenvectors of a system are degenerate, to complete the eigenbasis we introduce the Jordan chain equations \cite{Bronson69,Seyranian03}:

\begin{gather}
\begin{aligned}
S \ket{R_{EPD-L}} &= \lambda_{S}^{EPD-L} \ket{R_{EPD-L}}, \\
S \ket{J_{2}} &= \lambda_{S}^{EPD-L} \ket{J_{2}} + \ket{R_{EPD-L}}, \\
            & \vdotswithin{ = } \\
S \ket{J_{L}} &= \lambda_{S}^{EPD-L} \ket{J_{L}} + \ket{J_{L-1}},
\end{aligned}
\end{gather}

where $\lambda_{S}^{EPD-L}$, $\ket{R_{EPD-L}}$ are the degenerate eigenvalues and eigenvectors respectively, and $\ket{J_{L}}$ is the generalized eigenvector of rank $L$.

When injecting arbitrary signals into a three-port system at EPD-3 conditions, the input and output signals become: 

\begin{equation}
\begin{pmatrix}V_1^{\rm in}\\V_2^{\rm in}\\V_3^{\rm in} \end{pmatrix} = c_1 \ket{R_{EPD-3}} + c_2 \ket{J_{2}} + c_3 \ket{J_{3}}
\end{equation}

\begin{equation}
\begin{split}
\begin{pmatrix}V_1^{\rm out}\\V_2^{\rm out}\\V_3^{\rm out} \end{pmatrix} = \: &\lambda_S^{EPD-3} \left[c_1 \ket{R_{EPD-3}} + c_2 \ket{J_{2}} + c_3 \ket{J_{3}}\right] \\ & + c_2 \ket{R_{EPD-3}} + c_3 \ket{J_{2}}
\end{split}
\end{equation}

If the system is at both \rev{$det(S)=0+i0$} and EPD-3 conditions ($\lambda_S^{EPD-3}=0+i0$), the output signal becomes:

\begin{equation}
\begin{pmatrix}V_1^{\rm out}\\V_2^{\rm out}\\V_3^{\rm out} \end{pmatrix} = c_2 \ket{R_{EPD-3}} + c_3 \ket{J_{2}}.
\end{equation}

For all cases of signal injection into a three-port system described above, a robust splitting phenomenon related to \rev{$det(S)=0+i0$} conditions doesn't generically occur. To achieve a robust splitting, an additional constraint besides \rev{$det(S)=0+i0$} conditions is needed, such as setting $c_2$ or $c_3$ to be zero. This would involve exciting only two ports of the system to inject the signal, effectively reducing the dimensionality of the system. Nevertheless, any higher-order port system can exhibit robust splitting phenomena if only two ports are used for injecting and receiving signals, and \rev{RS} conditions are satisfied. 

All signal injection results above can be easily derived for higher dimensional scattering matrices ($M>3$), but robust splitting generically occurs only when injecting and receiving signals using two ports of the system.


\clearpage
\newpage

\bibliography{Bibliography}
\end{document}